\pdfoutput=1

\documentclass[preprint,pra,showpacs]{revtex4}%
\usepackage{amssymb}
\usepackage{amsfonts}
\usepackage{amsmath}
\usepackage{graphicx}%
\providecommand{\U}[1]{\protect\rule{.1in}{.1in}}

\begin{document}
\title[Quantum Time]{Time in Quantum Mechanics}
\author{Curt A. Moyer}
\affiliation{Department of Physics and Physical Oceanography, UNC Wilmington}
\email{moyerc@uncw.edu}

\begin{abstract}
The failure of conventional quantum theory to recognize time as an observable
and to admit time operators is addressed. Instead of focusing on the existence
of a time operator for a given Hamiltonian, we emphasize the role of the
Hamiltonian as the generator of translations in time to construct time states.
Taken together, these states constitute what we call a timeline, or quantum
history, that is adequate for the representation of any physical state of the
system. Such timelines appear to exist even for the semi-bounded and discrete
Hamiltonian systems ruled out by Pauli's theorem. However, the step from a
timeline to a valid time operator requires additional assumptions that are not
always met. Still, this approach illuminates the crucial issue surrounding the
construction of time operators, and establishes quantum histories as
legitimate alternatives to the familiar coordinate and momentum bases of
standard quantum theory.

\end{abstract}

\maketitle

\section{Introduction}

The treatment of time in quantum mechanics is one of the challenging open
questions in the foundations of quantum theory. On the one hand, time is the
parameter entering Schr\"{o}dinger's equation and measured by an external
laboratory clock. But time also plays the role of an \textit{observable} in
questions involving the occurrence of an event (e.g. when a nucleon decays, or
when a particle emerges from a potential barrier), and -- like every
observable -- should be represented in the theory by an operator whose
properties are predictors of the outcome of [event] time measurements made on
physical systems. Yet no time operators occur in ordinary quantum mechanics.
At its core, this is the \textit{quantum time problem}. As further testimony
to this conundrum, the uncertainty principle for time/energy is known to have
a different character than does the uncertainty principle for space/momentum.

An important landmark in the historical development of the subject is an early
`theorem' due to Pauli \cite{Pauli}. Pauli's argument essentially precludes
the existence of a self-adjoint time operator for systems where the spectrum
of the Hamiltonian is bounded, semi-bounded, or discrete, i.e., for most
systems of physical interest. Pauli concluded that \textquotedblleft\ldots the
introduction of an operator $\widehat{T}$ [time operator] must fundamentally
be abandoned\ldots\textquotedblright. While counterexamples to Pauli's theorem
are known, his assertion remains largely unquestioned and continues to be a
major influence shaping much of the present work in this area. For a
comprehensive, up-to-date review of this and related topics, see
\cite{TQM},\cite{TQM2}.

In this paper we advocate a different approach, one that emphasizes the
statistical distribution of event times -- not the time operator -- as the
primary construct. Essentially, we follow the program that regards time as a
POVM (positive, operator-valued measure) observable \cite{Srinivas}. Event
time distributions are calculated in the usual way from wave functions in the
time basis. We show that this time basis exists even for the semi-bounded and
discrete Hamiltonian systems ruled out by Pauli's theorem, and is adequate for
the representation of any physical state. However, the step from a time basis
to a valid time operator requires additional assumptions that are not always
met. Still, this approach illuminates the crucial issue surrounding the
construction of time operators and, at the same time (no pun intended),
establishes the time basis as a legitimate alternative to the familiar
coordinate and momentum bases of standard quantum theory.

The plan of the paper is as follows: In Sec. II we introduce the time basis
and establish the essential properties of its elements (the time states) for
virtually any physical system. Sec. III explores the relationship between time
states and time operators, and establishes existence criteria for the latter.
In Secs. IV and V we show how some familiar results for specific systems can
be recovered from the general theory advanced here, and in Sec. VI we obtain
new results applicable to the free particle in a three-dimensional space.
Results and conclusions are reported in Sec. VII. Throughout we adopt natural
units in which $\hbar=1$, a choice that leads to improved transparency by
simplifing numerous expressions.

\section{Quantum Histories: A Novel Basis Set}

We introduce basis states in the Hilbert space $|\,\tau\,\rangle$ labeled by a
real variable $\tau$ that we will call \textit{system time}. For a quantum
system described by the Hamiltonian operator $\widehat{H}$, these states are
defined by the requirement that $\widehat{H}$ be the generator of translations
among them. In particular, for any system time $\tau$ and every real number
$\alpha$\ we require%
\begin{equation}
|\,\tau+\alpha\rangle=\exp\left(  -i\widehat{H}\alpha\right)  |\,\tau
\,\rangle\label{basis property}%
\end{equation}
Since $\widehat{H}$ is assumed to be Hermitian, the transformation from
$|\,\tau\,\rangle$ to $|\,\tau+\alpha\rangle$ will be unitary (and therefore
norm-preserving). Eq.(\ref{basis property}) shows that if $|\,\tau\,\rangle$
is a member of this set then so too is $|\,\tau+\alpha\,\rangle$, implying
that system time $\tau$ extends continuously from the remote `past'
($\tau=-\infty$) to the distant `future' ($\tau=+\infty$). We refer to the set
$\left\{  |\,\tau\,\rangle:-\infty\leq\tau\leq\infty\right\}  $ as a
\textit{timeline} or \textit{quantum history}.

Eq.(\ref{basis property}) is reminiscent of the propagation of quantum states,
which evolve according to $|\,\psi(t)\,\rangle=\exp\left(  -i\widehat{H}%
t\right)  |\,\psi(0)\,\rangle$. It follows that the dynamical wave function in
the time basis, $\langle\,\tau\,|\,\psi(t)\,\rangle$, obeys%
\begin{equation}
\langle\,\tau\,|\,\psi(t)\,\rangle\,=\langle\,\tau-t\,|\,\psi(0)\,\rangle,
\label{covariance}%
\end{equation}
a property known in the quantum-time literature as \textit{covariance}. To
appreciate its significance, recall that $\left\vert \langle\,\tau
\,|\,\psi(t)\,\rangle\right\vert ^{2}$\ is essentially the probability that a
given [system] time $\tau$ will be associated with some measurement after a
[laboratory] time $t$ has passed; covariance ensures that the same probability
will be obtained for the initial state at the earlier [system] time $\tau-t$.
This is time-translation invariance, widely recognized as an essential feature
that must be reproduced by any statistical distribution of time observables
\cite{Kijowski}.

We advance the conjecture that \textit{a quantum history exists for every
system, with elements (the time states) sufficiently numerous to span the
Hilbert space of physical states}. The completeness of this basis is expressed
in the abstract by the following resolution of the identity:%
\begin{equation}
1=%
{\displaystyle\int_{-\infty}^{\infty}}
|\,\tau\,\rangle\langle\,\tau\,|\,d\tau\label{closure rule}%
\end{equation}
Specifically, we insist that for any two normalizable states $|\,\psi
\,\rangle$\ and $|\,\varphi\,\rangle$, we must be able to write%
\begin{equation}
\langle\varphi\,|\,\psi\,\rangle=%
{\displaystyle\int\limits_{-\infty}^{\infty}}
\,\langle\varphi\,|\,\tau\,\rangle\langle\,\tau\,|\,\psi\,\rangle
\,d\tau\label{closure-L2}%
\end{equation}
The time states also are orthogonal, at least in a `weak' sense consistent
with closure. More precisely, if $|\,\varphi\,\rangle$ in Eq.(\ref{closure-L2}%
) can be replaced by a time state, we get%
\begin{equation}
\langle\,\tau\,|\,\psi\,\rangle=%
{\displaystyle\int_{-\infty}^{\infty}}
\langle\,\tau\,|\,\tau^{\prime}\,\rangle\langle\,\tau^{\prime}\,|\,\psi
\,\rangle\,d\tau^{\prime} \label{weak-orthogonality}%
\end{equation}
Eq.(\ref{weak-orthogonality}) expresses quantitatively the notion of `weak'
orthogonality; it differs from standard usage (`strong' orthogonality) by
uniquely specifying the domain of integration and restricting $|\,\psi
\,\rangle$ to be a normalizable state. But owing to the continuous nature of
time, the time states $|\,\tau\rangle$ are \textit{not} normalizable, so
Eq.(\ref{closure-L2}) can be satisfied even when Eq.(\ref{weak-orthogonality})
is not. It follows that `weak' orthogonality is a stronger condition than
closure, and subject to separate verification.

Quantum histories are intimately related to spectral structure, and derivable
from it. Indeed, we regard as axiomatic the premise that the eigenstates of
$\widehat{H}$, say $|\,E\,\rangle$, also span the space of physically
realizable states to form the \textit{spectral basis} $\left\{  |\,E\,\rangle
:\forall E\right\}  $. Since $\widehat{H}$ is by assumption Hermitian, the
elements of the spectral basis can always be made mutually orthogonal (in the
`strong' sense), and normalized so as to satisfy a closure rule akin to
Eq.(\ref{closure rule}). Now Eq.(\ref{basis property}) dictates that the
timeline--spectral transformation is characterized by functions $\langle
\,\tau\,|\,E\,\rangle$ such that $\langle\,\tau+\alpha\,|\,E\,\rangle
=\exp\left(  iE\alpha\right)  \langle\,\tau\,|\,E\,\rangle$; equivalently
(with $\alpha\rightarrow\alpha-\tau$),%
\begin{equation}
\exp\left(  -iE\tau\right)  \langle\,\tau\,|\,E\,\rangle=c_{E}
\label{timeline-spectral link}%
\end{equation}
This form is imposed by covariance. The constant on the right, while
unspecified, is manifestly independent of $\tau$. This enables us to write%
\begin{equation}
\langle\,\tau\,|\,\psi\,\rangle\equiv\sum\limits_{E}c_{E}\exp\left(
iE\tau\right)  \langle\,E\,|\,\psi\,\rangle\, \label{history-general}%
\end{equation}
The sum in this equation is symbolic, translating into an ordinary sum over
any discrete levels together with an integral over the continuum. \textit{A
timeline exists if for every physical state }$|\,\psi\,\rangle$\textit{ the
transform of Eq.(\ref{history-general}) maps the set }$\left\{  \langle
\,E\,|\,\psi\,\rangle:\forall E\right\}  $\textit{ into a square-integrable
function }$\langle\,\tau\,|\,\psi\,\rangle$ on $-\infty<\tau<\infty$. The
remainder of this section is devoted to confirming the existence of these
timelines and verifying their essential properties (covariance, completeness,
`weak' orthogonality) for a wide variety of spectra. More generally, we
contend that timelines \textit{always} can be constructed with the properties
expressed by Eqs.(\ref{basis property}) and (\ref{closure rule}) above. In
this sense the Hamiltonian $\widehat{H}$ can be said to define its own time --
an intrinsic [system] time $\tau$ -- that is quite distinct from the
laboratory time $t$ appearing in the Schr\"{o}dinger equation.

\subsection{Accessible States Model: Non-degenerate Spectra with a Recurrence
Time}

We consider here spectra consisting exclusively of discrete, non-degenerate
levels. The spacing between adjacent levels might be arbitrarily small
(approximating a continuum), but all level separations are assumed to be
non-zero. The levels are therefore countable; we ascribe to them energies
$E_{j}:j=1,2,\ldots$ ordered by increasing energy, and label as $|\,E_{j}%
\,\rangle$ the state with energy $E_{j}$. Consistent with a Hamiltonian
operator that is Hermitian, we will take these states to be mutually
orthogonal and normalized (to unity). Generally, there will be an infinite
number of such states that together span the Hilbert space of physical states.
In the \textit{accessible states model} the first $N$ of these states are
deemed sufficient to represent a given physical system subject to the
available interactions. By increasing their number, we eventually include all
states that are accessible from one another by some series of physical
interactions. In this way we are left at every stage with a \textit{finite}
[$N$-dimensional] Hilbert space spanned by a spectral basis whose elements are
orthonormal. Admittedly, $N$ can be very large and is somewhat ill-defined; in
consequence, results -- to be useful -- will have to be reported in a way that
makes clear how we transition to the $N\rightarrow\infty$ limit. Lastly, for
every $N$ we assume the existence of a \textit{recurrence time}, a reference
to the smallest duration over which an arbitrary initial state regains its
initial form. Recurrence has implications for spectral structure, without
being unduly restrictive. Indeed, we believe the model just outlined can serve
as a basic template for any realistic spectrum. Our approach is essentially
that taken by Pegg \cite{Pegg} in his search for an operator conjugate to the
Hamiltonian of periodic systems. What follows amounts to a restatement of
those basic results and their extension to aperiodic systems, where the
recurrence time becomes arbitrarily large.

In keeping with Eq.(\ref{timeline-spectral link}), the spectral-time
transformation in the accessible states model is specified by the functions%
\[
\langle\,\tau\,|\,E_{j}\,\rangle\equiv c_{j}\exp\left(  iE_{j}\tau\right)
\qquad j=1,2,\ldots,N
\]
The coefficients $c_{j}$\ are chosen to ensure closure of the resulting time
states; in turn, this requires for all $k\neq j$%
\[
0=%
{\displaystyle\int}
\langle\,E_{k}\,|\,\tau\,\rangle\langle\,\tau\,|\,E_{j}\,\rangle\,d\tau
=c_{k}^{\ast}c_{j}%
{\displaystyle\int}
\exp\left(  iE_{j}\tau-iE_{k}\tau\right)  \,d\tau,
\]
i.e., the functions $\left\{  \exp\left(  iE_{j}\tau\right)  :\forall
j\right\}  $ must constitute an orthogonal set over the domain of the time
label $\tau$. This domain derives from the recurrence time (also known as the
\textit{revival time}) denoted here by $\tau_{rev}$, a reference to the
smallest duration over which an arbitrary initial state $|\,\psi\,\rangle$
regains its initial form, up to an overall [physically insignificant] phase.
Since state evolution is governed by the system Hamiltonian, recurrence has
consequences for the energy spectrum: specifically, a revival time implies the
existence of a smallest integer $n_{j}$ for every energy level $E_{j}$ such
that%
\begin{equation}
E_{j}\tau_{rev}=2\pi n_{j}+\theta\label{revivals}%
\end{equation}
To take a familiar example, let's assume the level distribution is uniform
with spacing $\Delta\epsilon$, so that $E_{j}=E_{0}+j\,\Delta\epsilon,\quad
j=0,1,2,\ldots$ (uniform level spacing is the hallmark of the harmonic
oscillator). For this case $n_{j}=j$, $\theta=E_{0}\tau_{rev}$, and the
recurrence time is just $\tau_{rev}=2\pi/\Delta\epsilon$. Another well-known
example is the infinite square well, for which $E_{j}=j^{2}\,E_{1},\quad
j=1,2,\ldots$ and $E_{1}$ is the energy of the ground state. In this instance
we have $n_{j}=j^{2}$, $\theta=0$, and a recurrence time $\tau_{rev}%
=2\pi/E_{1}$. One consequence of Eq.(\ref{revivals}) is that the energy
spectrum is commensurate, meaning that the ratio of any two levels -- after
adjusting for a possible global offset -- is a rational number. That
commensurability is a necessary but not sufficient condition for recurrence is
illustrated by the discrete spectrum of hydrogen: $E_{j}=-\,E_{1}/j^{2},\quad
j=1,2,\ldots$ with $-E_{1}$ the ground state energy. Here commensurability is
evident (with zero offset), but there is no [finite] revival time unless the
spectrum is truncated, say at $j=N$. This recurrence time clearly grows with
increasing $N$ (the largest period is $2\pi/\left\vert E_{N}\right\vert $),
and becomes infinite when all levels are included.

With a revival time we should be able to limit the domain of $\tau$ to a
single recurrence cycle; indeed, a straightforward calculation with the help
of Eq.(\ref{revivals}) reveals that the functions $\left\{  \exp\left(
iE_{j}\tau\right)  :j=0,1,2,\ldots\right\}  $ are in fact orthogonal over any
one cycle, thereby rendering the spectral--time transformation essentially a
Fourier series. In fact, with the understanding that $\langle\,E_{n}%
\,|\,\psi\,\rangle$ is zero unless $n$ is a member of the set $\left\{
n_{j}\right\}  $ defined by Eq.(\ref{revivals}), Eq.(\ref{history-general})
can be recast as%
\[
\langle\,\tau\,|\,\psi\,\rangle=\exp\left(  i\theta\tau/\tau_{rev}\right)
{\displaystyle\sum_{n}}
c_{n}\exp\left(  i2\pi n\tau/\tau_{rev}\right)  \,\langle\,E_{n}%
\,|\,\psi\,\rangle
\]
But for the multiplicative phase factor, this \textit{is} a Fourier series for
functions that are periodic with period $\tau_{rev}$.

At this point we could simply invoke a general result of Fourier analysis\ to
argue that the above series converges [in the norm] to define a viable
timeline function $\langle\,\tau\,|\,\psi\,\rangle$ for any normalizable state
$|\,\psi\,\rangle$. However, it is far more illuminating to investigate in
some detail how this actually comes about. To that end, we begin with an
important observation reinforced by our earlier example of the discrete
hydrogen spectrum: \textit{for a system with a finite number of energy levels
-- no matter how large -- a revival time always exists in practice}. To see
this, consider the $N-1$ `gaps' $\Delta E_{j}\equiv E_{j+1}-E_{j}$ separating
adjacent discrete levels, and denote the smallest of these by $\Delta E_{\min
}$. If the gap ratio $\Delta E_{j}/\Delta E_{\min}$ is a rational number, say
$p_{j}/q_{j}$, then a revival time exists and is given by $\tau_{rev}%
=2\pi\cdot\left(  \Pi_{j}q_{j}\right)  /\Delta E_{\min}$. Since any irrational
number can be approximated to arbitrary precision by a rational one, a revival
time `almost' always exists (though it may far exceed the natural period
$2\pi/\left\vert E_{j}\right\vert $ associated with any one energy).

With $N$ orthogonal stationary states there can be no more than an identical
number of linearly independent time states, i.e., the time basis $\left\{
|\,\tau\,\rangle:\tau_{0}\leq\tau<\tau_{0}+\tau_{rev}\right\}  $ must be
vastly \textit{overcomplete}. This suggests that closure might be achieved
with just $N$ discrete time states, properly chosen. To verify this, we select
from the time domain $\left[  \tau_{0},\tau_{0}+\tau_{rev}\right]  $ a uniform
mesh of $N$ points $\left\{  \tau_{p}:p=1,2,\ldots N\right\}  $ and evaluate%
\[%
{\displaystyle\sum_{p=1}^{N}}
\langle E_{k}\,|\,\tau_{p}\,\rangle\langle\,\tau_{p}\,|\,E_{j}\,\rangle
=c_{k}^{\ast}c_{j}%
{\displaystyle\sum_{p=1}^{N}}
\exp\left[  i\left(  E_{j}-E_{k}\right)  p\Delta\tau\right]
\]
Each term in the sum on the right can be represented by a unit-amplitude
phasor. The phasor sum is obtained by adding successive phasors tail-to-tip,
resulting in an $N$-sided regular convex polygon with each exterior angle
(angle between successive phasors) equal to $\left(  E_{j}-E_{k}\right)
\Delta\tau$. But since $N\Delta\tau=\tau_{rev}$, the cumulative exterior angle
is always an integer multiple of $2\pi$ ($\left(  E_{j}-E_{k}\right)
\tau_{rev}=2\pi\left(  n_{j}-n_{k}\right)  $ from Eq.(\ref{revivals})),
implying that said polygon invariably is \textit{closed}. Thus the sum
vanishes for all $E_{k}\neq E_{j}$. (For $E_{k}=E_{j}$ the value is simply
$N$, the number of terms in the sum.) Since $|\,E_{j}\,\rangle$ and
$|\,E_{k}\,\rangle$ refer to any pair of stationary states (and these are
complete by hypothesis),\ the preceding results imply the operator
equivalence
\[
\frac{1}{N}%
{\displaystyle\sum_{p=1}^{N}}
\,|\,\tau_{p}\,\rangle\langle\,\tau_{p}\,|\,=%
{\displaystyle\sum_{j=1}^{N}}
\,\left\vert c_{j}\right\vert ^{2}|\,E_{j}\,\rangle\langle\,E_{j}\,|
\]
\ It follows that these $N$ consecutive time states by themselves span this
$N$-dimensional space provided $\left\vert c_{j}\right\vert >0$ for every $j$
(the only vector orthogonal to all is the null element). Closure equivalence
with the spectral basis further requires all coefficients $c_{j}$ to have unit
magnitude; since any phase can be absorbed into the stationary state
$|\,E_{j}\,\rangle$, we will simply take $c_{j}=1$.

The time basis $\left\{  |\,\tau_{p}\,\rangle:p=1,2,\ldots N\right\}  $ has
discrete elements, yet $\tau$ supposedly is a continuous variable. We resolve
this apparent contradiction by noting that the argument leading to this
discrete time basis is unaffected if $N$ is scaled by any integer and
$\Delta\tau$ is reduced by the same factor (leaving $N\Delta\tau=\tau_{rev}$
unchanged). Thus, every set of time states%
\begin{equation}
\left\{  |\,\tau_{p}^{\left(  n\right)  }\,\rangle:p=1,2,\ldots nN:\Delta
\tau^{(n)}=\tau_{rev}/nN,\;n=1,2,\ldots\right\}  \label{basis-discrete}%
\end{equation}
is complete in this $N$-dimensional space, but all except the primitive one
($n=1$) is \textit{over}complete. Nonetheless, this observation allows us to
write%
\begin{equation}%
{\displaystyle\sum_{j=1}^{N}}
\,|\,E_{j}\,\rangle\langle\,E_{j}\,|\,=\frac{1}{\tau_{rev}}%
{\displaystyle\sum_{\tau=\tau_{0}}^{\tau_{0}+\tau_{rev}}}
\,\,|\,\tau_{p}^{(n)}\,\rangle\langle\,\tau_{p}^{(n)}\,|\,\Delta\tau
^{(n)}\longrightarrow\frac{1}{\tau_{rev}}\int_{\tau_{0}}^{\tau_{0}+\tau_{rev}%
}|\,\tau\,\rangle\langle\,\tau\,|\,d\tau\label{closure-discrete}%
\end{equation}
(Passage to the continuum limit presumes that $\tau$ has no granularity, even
on the smallest scale.) Scaling the time states by $\sqrt{1/\tau_{rev}}$ then
results in an indenumerable time basis that enjoys closure equivalence with
the spectral basis in this $N$-dimensional space. Explicitly, the squared norm
of any element $|\,\psi\,\rangle$ can be expressed as%
\begin{equation}
\langle\,\psi\,|\,\psi\,\rangle=%
{\displaystyle\sum_{j=1}^{N}}
\,\left\vert \langle\,E_{j}\,|\,\psi\,\rangle\right\vert ^{2}=\int_{\tau_{0}%
}^{\tau_{0}+\tau_{rev}}\left\vert \langle\,\tau\,|\,\psi\,\rangle\right\vert
^{2}\,d\tau\label{closure equivalence}%
\end{equation}
with%
\begin{equation}
\langle\,\tau\,|\,\psi\,\rangle=\frac{1}{\sqrt{\tau_{rev}}}%
{\displaystyle\sum_{j=1}^{N}}
\exp\left(  iE_{j}\tau\right)  \,\langle\,E_{j}\,|\,\psi\,\rangle
\label{history-discrete}%
\end{equation}

While it could be argued that $N$ always has an upper bound in practice, it is
typically quite large and known only imprecisely. Convergence of these results
as $N\rightarrow\infty$ therefore is crucial to a viable theory of timelines.
Now for every $N$ no matter how large, Eq.(\ref{closure-discrete}) establishes
closure equivalence of the time basis with the spectral basis, and leads
directly to the \textit{Plancherel identity} \cite{Olver} expressed by
Eq.(\ref{closure equivalence}). But by definition, the norm of a normalizable
state remains finite even in the limit $N\rightarrow\infty$, so the Plancherel
identity ensures the existence of the square-integrable function
$\langle\,\tau\,|\,\psi\,\rangle$ on $\left[  \tau_{0},\tau_{0}+\tau
_{rev}\right]  $ in this same limit. We conclude that
Eq.(\ref{history-discrete}) maps the spectral components $\langle
\,E_{j}\,|\,\psi\,\rangle$ into square-integrable functions $\langle
\,\tau\,|\,\psi\,\rangle$ in the time basis, as required for a timeline.\ This
establishes convergence [in the norm] for the timeline wave function
$\langle\,\tau\,|\,\psi\,\rangle$.

Our final task is to write the transformation law of
Eq.(\ref{history-discrete}) in a form that is useful for calculation in the
large $N$ limit. The way we proceed depends on how the recurrence time varies
with $N$. As more states are included in the model, all levels might remain
isolated no matter how numerous they become ($\tau_{rev}$ saturates at a
finite value); alternatively, some levels may cluster to form a
quasi-continuum ($\tau_{rev}\rightarrow\infty$), merging into a true continuum
as $N\rightarrow\infty$. Anticipating a mix of the two, we write the
quasi-continuum contributions to $\langle\,\tau\,|\,\psi\,\rangle$ in terms of
the characteristic energy $\Delta E\equiv2\pi/\tau_{rev}$ associated with the
recurrence time $\tau_{rev}$:%
\[
\frac{1}{\sqrt{\tau_{rev}}}%
{\displaystyle\sum_{\text{continuum}}}
\left(  \ldots\right)  =\frac{1}{\sqrt{2\pi}}%
{\displaystyle\sum_{\text{continuum}}}
\exp\left(  iE_{j}\tau\right)  \,\left[  \frac{\langle\,E_{j}\,|\,\psi
\,\rangle}{\sqrt{\Delta E}}\right]  \Delta E
\]
The significance of the bracketed term can be appreciated by comparing
discrete and continuum contributions to the squared norm of the state
$|\,\psi\,\rangle$:%
\[
\langle\,\psi\,|\,\psi\,\rangle=%
{\displaystyle\sum_{\text{discrete}}}
\,\left\vert \langle\,E_{j}\,|\,\psi\,\rangle\right\vert ^{2}+%
{\displaystyle\sum_{\text{continuum}}}
\,\left\vert \frac{\langle\,E_{j}\,|\,\psi\,\rangle}{\sqrt{\Delta E}%
}\right\vert ^{2}\Delta E
\]
The replacement%
\[
\frac{\langle\,E_{j}\,|\,\psi\,\rangle}{\sqrt{\Delta E}}\rightarrow
\langle\,E_{j}\,|\,\psi\,\rangle
\]
amounts to a renormalization of the quasi-continuum wave function -- known as
\textit{energy normalization} \cite{Merzbacher} -- such that the [energy-]
integrated density of the new function carries the same weight as does an
isolated state. Replacing sums over the quasi-continuum with integrals becomes
exact in the large $N$ limit (with the inclusion of additional levels,
$\tau_{rev}\rightarrow\infty$ and $\Delta E\rightarrow0$). In this way, we
arrive at a form of the transformation law that lends itself to computation as
$N\rightarrow\infty$:%
\begin{equation}
\langle\,\tau\,|\,\psi\,\rangle=\frac{1}{\sqrt{\tau_{rev}}}%
{\displaystyle\sum_{\text{discrete}}}
\exp\left(  iE_{j}\tau\right)  \,\langle\,E_{j}\,|\,\psi\,\rangle+\frac
{1}{\sqrt{2\pi}}\int_{E_{\min}}^{E_{\max}}\exp\left(  iE\tau\right)
\langle\,E\,|\,\psi\,\rangle\,dE \label{history-mixed}%
\end{equation}
Eq.(\ref{history-mixed})\ assumes that all discrete stationary states are
normalized to unity ($\langle\,E_{j}\,|\,E_{k}\,\rangle=\delta_{jk}$), and all
[quasi-]continuum elements are energy-normalized ($\langle\,E\,|\,E^{\prime
}\,\rangle=\delta\left(  E\,-\,E^{\prime}\right)  $). Notice that the
continuum contribution references $N$ only indirectly through the spectral
bounds $E_{\min}$, $E_{\max}$ \cite{eMax}. Furthermore, whenever a continuum
is present (aperiodic systems, for which $\tau_{rev}\rightarrow\infty$), it
makes the dominant contribution to $\langle\,\tau\,|\,\psi\,\rangle$ for any
fixed value of $\tau$. To be fair, there is one limitation: $\tau$ cannot be
so large that $\exp\left(  iE\tau\right)  $ varies appreciably over the step
size $\Delta E=2\pi/\tau_{rev}$; this means the integral approximation to the
quasi-continuum fails for $\tau\simeq\tau_{rev}$. In a similar vein, the
discrete terms can never simply be dropped from Eq.(\ref{history-mixed}), as
they make the largest contribution to $\langle\,\tau\,|\,\psi\,\rangle$ in the
asymptotic regime.

Although complete, the time basis typically includes non-orthogonal elements.
In the absence of a continuum, and with uniform level spacing for the discrete
terms, it is not difficult to show that the minimal time basis is composed of
$N$ \textit{mutually orthogonal} states. But otherwise the existence of even
one orthogonal pair is not guaranteed. By contrast, `weak' orthogonality is
the rule in this $N$-dimensional space. This can be verified rigorously by
using closure of the [discrete] time basis in Eq.(\ref{basis-discrete}) to
write%
\[
\langle\,\tau_{q}^{(n)}\,|\,\psi\,\rangle=\frac{1}{\tau_{rev}}%
{\displaystyle\sum_{\tau=\tau_{0}}^{\tau_{0}+\tau_{rev}}}
\,\,\langle\,\tau_{q}^{(n)}\,|\,\tau_{p}^{(n)}\,\rangle\langle\,\tau_{p}%
^{(n)}\,|\,\psi\,\rangle\,\Delta\tau^{(n)}%
\]
Scaling the time states by $\sqrt{1/\tau_{rev}}$ and passing to the continuum
limit then gives%
\begin{equation}
\langle\,\tau\,|\,\psi\,\rangle=\int_{\tau_{0}}^{\tau_{0}+\tau_{rev}}%
\langle\,\tau\,|\,\tau^{\prime}\,\rangle\langle\,\tau^{\prime}\,|\,\psi
\,\rangle\,d\tau^{\prime}, \label{weak-ortho-discrete}%
\end{equation}
with $\langle\,\tau\,|\,\psi\,\rangle$\ again calculated from
Eq.(\ref{history-mixed}). Since $N$ is not referenced explicitly here, we
conclude from Eq.(\ref{weak-ortho-discrete}) that `weak' orthogonality
persists in the limit as $N\rightarrow\infty$ at every value $\tau$ where the
timeline wave function $\langle\,\tau\,|\,\psi\,\rangle$ converges.

\subsection{The Treatment of Degeneracy}

Degenerate states require labels in addition to the energy to distinguish
them. These extra labels derive from the underlying symmetry that is the root
of all degeneracy. Thus, a central potential gives rise to a
rotationally-invariant Hamiltonian and this, in turn, implies that the
Hamiltonian operator commutes with angular momentum (the generator of
rotations). In such cases, the energy label is supplemented with orbital and
magnetic quantum numbers specifying the particle angular momentum. The larger
point is simply this: with the additional labels comes again an unambiguous
identification of the spectral states, and we write $|\,E_{j}\,\rangle
\rightarrow|\,E_{j}\,,\sigma\,\rangle$, where $\sigma$ is a collective label
that symbolizes \textit{all} additional quantum numbers needed to identify the
spectral basis element with a given energy. With this simple modification, the
arguments of the preceding section remain intact. Quantum histories can be
constructed as before, but now are indexed by the same `good' quantum numbers
that characterize the spectrum. That is, in the face of degeneracy we have not
one -- but multiple -- timelines, and we write $|\,\tau\,\rangle
\rightarrow|\,\tau,\sigma\,\rangle$. Notice that any two time states belonging
to \textit{distinct} timelines will be orthogonal (in the `strong' sense)%
\begin{equation}
\langle\,\tau,\sigma\,|\,\tau^{\prime},\sigma^{\prime}\,\rangle\propto
\delta_{\sigma,\sigma^{\prime}},
\end{equation}
and -- of course -- \textit{all} timelines must be included to span the entire
Hilbert space, so that Eq.(\ref{closure rule}) becomes%
\begin{equation}
1=\sum_{\sigma}%
{\displaystyle\int_{-\infty}^{\infty}}
|\,\tau,\sigma\,\rangle\langle\tau,\sigma\,|\,d\tau
\end{equation}

Oftentimes there is more than one way to select good quantum numbers.
Following along with our earlier example, states having different magnetic
quantum numbers may be combined to describe orbitals with highly directional
characteristics that are important in chemical bonding. The crucial point here
is that the new states are related to the old by a \textit{unitary}
transformation (a `rotation') in the subspace spanned by the degenerate
states. In turn, the `rotated' stationary states gives rise to new time
states, related to the old as%
\begin{equation}
|\,\tau^{\left(  r\right)  }\,\rangle=\sum_{\sigma}U_{r\sigma}|\,\tau
,\sigma\,\rangle, \label{subspace-rotation}%
\end{equation}
where $U_{r\sigma}$ are elements of the \textit{same} unitary matrix that
characterizes the subspace `rotation' (cf. Eq.(\ref{history-general})).
Furthermore, unitarity ensures that the projector onto every degenerate
subspace is representation-independent:%
\begin{align}
\sum_{r}|\,\tau^{\left(  r\right)  }\,\rangle\langle\,\tau^{(r)}\,|\,  &
=\sum_{\sigma,\sigma^{\prime}}\left(  \sum_{r}U_{r\sigma}U_{r\sigma^{\prime}%
}^{\ast}\right)  |\,\tau,\sigma\,\rangle\langle\,\tau,\sigma^{\prime
}\,|\nonumber\\
\,  &  =\sum_{\sigma,\sigma^{\prime}}\left(  U^{\dag}U\right)  _{\sigma
^{\prime}\sigma}|\,\tau,\sigma\,\rangle\langle\,\tau,\sigma^{\prime}%
\,|\,=\sum_{\sigma}|\,\tau,\sigma\,\rangle\langle\,\tau,\sigma\,|
\label{subspace-projector}%
\end{align}
As we shall soon see, Eq.(\ref{subspace-projector}) has important
ramifications for the statistics of [event] time observables whenever
degeneracy is present.

One final observation: degenerate or not, the timelines of
Eq.(\ref{history-mixed}) are \textit{not} unique, inasmuch as they can be
altered by an (energy-dependent) phase adjustment to the eigenstates of
$\widehat{H}$. Consider the replacement $|\,E\,\rangle\rightarrow\exp\left(
i\theta_{E}\right)  |\,E\,\rangle$. If $\theta_{E}$ is proportional to $E$,
say $\theta_{E}=\tau_{0}E$, then $|\,\tau\,\rangle\rightarrow|\,\tau-\tau
_{0}\,\rangle$, i.e., a linear (with energy) phase adjustment to the
stationary states shifts the origin of system time, underscoring the notion
that only \textit{durations} in system time can have measureable consequences.
Other, more complicated phase adjustment schemes can be contemplated, with the
time states always related by a suitable unitary transformation. Some of these
have clear physical significance, as later examples will show.

\section{Interpreting Time States, and an Operator for Local Time}

The time states of Sec. II can be used to formally construct a time operator;
for non-degenerate spectra,%
\begin{equation}
\widehat{T}=%
{\displaystyle\int\limits_{\tau_{0}}^{\tau_{0}+\tau_{rev}}}
|\,\tau\,\rangle\tau\langle\,\tau\,|\,d\tau\label{time operator}%
\end{equation}
(Degeneracy requires the replacement $|\,\tau\,\rangle\langle\,\tau
\,|\,\rightarrow$\ $\sum_{\sigma}|\,\tau,\sigma\,\rangle\langle\,\tau
,\sigma\,|$ in this and subsequent expressions.) At a minimum, the existence
of $\widehat{T}$ demands that matrix elements of Eq.(\ref{time operator})
taken between any two normalizable states $|\,\varphi\,\rangle$ and
$|\,\psi\,\rangle$ be well-defined, i.e., $\left\vert \langle\varphi
\,|\widehat{T\,}|\,\psi\,\rangle\right\vert <\infty$. Since the arguments of
Sec. II show that the timeline functions $\langle\,\tau\,|\,\psi\,\rangle$ are
square-integrable, this criterion clearly is met for all finite values of
$\tau_{0}$ and $\tau_{rev}$. The issue assumes greater importance if
$\tau_{rev}$ becomes arbitrarily large (aperiodic systems); we will return to
this point later.

Because time states are generally non-orthogonal and overcomplete, the
significance of this time operator is not clear. Indeed, the states
$|\,\tau\,\rangle$ cannot be eigenstates of the operator $\widehat{T}$, which
is Hermitian (as evidenced by its matrix elements in the spectral basis).
Nonetheless, the time basis constitutes a resolution of the identity, so the
timeline wave function may admit a probability interpretation along
conventional lines. For any state $|\,\psi\,\rangle$ (cf.
Eq.(\ref{closure equivalence}))%
\[
\langle\psi\,|\,\psi\,\rangle=%
{\displaystyle\int\limits_{\tau_{0}}^{\tau_{0}+\tau_{rev}}}
\left\vert \langle\,\tau\,|\,\psi\,\rangle\right\vert ^{2}\,d\tau,
\]
suggesting that $\left\vert \langle\,\tau\,|\,\psi\,\rangle\right\vert
^{2}d\tau$ -- besides being positive definite -- is additive for disjoint sets
and sums to unity for any properly normalized state, all attributes of a
bona-fide probability distribution. In turn, this begs the question: if
$\left\vert \langle\,\tau\,|\,\psi\,\rangle\right\vert ^{2}\,d\tau$ is a
probability, to what does this probability refer? The answer is found in the
notion of probability-operator measures, which asserts that the states
$|\,\tau\,\rangle$ provide a \textit{positive operator-valued measure (POVM)}
for the system time $\tau$; specifically, $\left\vert \langle\,\tau
\,|\,\psi\,\rangle\right\vert ^{2}\,d\tau$ represents the probability that a
suitable measuring instrument will return a result for system time between
$\tau$ and $\tau+d\tau$ \cite{Helstrom}. How one actually performs such a
measurement is an interesting question in its own right, and will not be
examined here (but see Ref.\cite{Hegerfeldt}).

The utility of the time operator defined by Eq.(\ref{time operator}) is that
its expectation value in any normalized state $|\,\psi\,\rangle$ furnishes the
average system time, or the `first moment' of this POVM:%
\begin{equation}
\tau_{avg}=%
{\displaystyle\int\limits_{\tau_{0}}^{\tau_{0}+\tau_{rev}}}
\tau\left\vert \langle\,\tau\,|\,\psi\,\rangle\right\vert ^{2}\,d\tau
=\langle\,\psi\,|\widehat{T}|\,\psi\,\rangle
\end{equation}
For any finite value of $\tau_{rev}$ the `second moment' of this POVM also
exists, and implies that the function $\langle\,\tau\,|\,\varphi
\,\rangle\equiv\tau\langle\,\tau\,|\,\psi\,\rangle$ is normalizable on
$\left[  \tau_{0},\tau_{0}+\tau_{rev}\right]  $. Applying `weak' orthogonality
to $\langle\,\tau\,|\,\varphi\,\rangle$ gives (cf.
Eq.(\ref{weak-ortho-discrete}))%
\begin{equation}
\tau\langle\,\tau\,|\,\psi\,\rangle=%
{\displaystyle\int\limits_{\tau_{0}}^{\tau_{0}+\tau_{rev}}}
\langle\,\tau\,|\,\tau^{\prime}\,\rangle\tau^{\prime}\langle\,\tau^{\prime
}\,|\,\psi\,\rangle\,d\tau^{\prime}=\langle\,\tau\,|\widehat{T}|\,\psi
\,\rangle\label{weak-eigenstate}%
\end{equation}
Since Eq.(\ref{weak-eigenstate}) holds for every normalizable state
$|\,\psi\,\rangle$, $|\,\tau\,\rangle$ is sometimes said to be a `weak'
eigenstate of $\widehat{T}$ \cite{Giannitrapani}. Being a `weak' eigenstate
has important consequences; for one, it allows us to write the `second moment'
of this POVM as%
\[%
{\displaystyle\int\limits_{\tau_{0}}^{\tau_{0}+\tau_{rev}}}
\tau^{2}\left\vert \langle\,\tau\,|\,\psi\,\rangle\right\vert ^{2}\,d\tau=%
{\displaystyle\int\limits_{\tau_{0}}^{\tau_{0}+\tau_{rev}}}
\langle\,\psi\,|\widehat{T}|\,\tau\,\rangle\langle\,\tau\,|\widehat{T}%
|\,\psi\,\rangle\,d\tau=\langle\,\psi\,|\widehat{T}^{2}|\,\psi\,\rangle,
\]
In turn, the variance of the time distribution can be expressed in the form%
\begin{equation}
\left(  \Delta\tau\right)  ^{2}\equiv%
{\displaystyle\int\limits_{\tau_{0}}^{\tau_{0}+\tau_{rev}}}
\left(  \tau-\tau_{avg}\right)  ^{2}\left\vert \langle\,\tau\,|\,\psi
\,\rangle\right\vert ^{2}\,d\tau=\langle\,\psi\,|\widehat{T}^{2}%
|\,\psi\,\rangle-\langle\,\psi\,|\widehat{T}|\,\psi\,\rangle^{2},
\end{equation}
which leads in the usual way to an uncertainty principle for [system] time and
energy:%
\begin{equation}
\Delta\tau\cdot\Delta E\geq\frac{1}{2}\left\vert \langle\,\psi\,|\left[
\widehat{T},\widehat{H}\right]  |\,\psi\,\rangle\right\vert
\label{uncertainty product}%
\end{equation}
For a \textit{canonical time operator}, $\left[  \widehat{T},\widehat{H}%
\right]  =i$, and we recover the familar uncertainty relation $\Delta\tau
\cdot\Delta E\geq1/2$, but as will soon become clear, canonical time operators
are not to be expected in periodic systems. On a related note,
Eq.(\ref{uncertainty product}) does \textit{not} extend to aperiodic systems
unless the variance $\left(  \Delta\tau\right)  ^{2}$ can be shown to exist in
the aperiodic limit $\tau_{rev}\rightarrow\infty$.

Eq.(\ref{time operator}) actually defines a family of related time operators
$\widehat{T}(\tau_{0})$ whose members are parametrically dependent on the
continuous variable $\tau_{0}$. The relationship between family members is
readily demonstrated. On the one hand, we appeal to Eq.(\ref{basis property})
to write for any real number $\alpha$%
\begin{align*}
\widehat{T}\left(  \tau_{0}+\alpha\right)   &  =%
{\displaystyle\int\limits_{\tau_{0}+\alpha}^{\tau_{0}+\alpha+\tau_{rev}}}
|\,\tau\,\rangle\tau\langle\,\tau\,|\,d\tau=%
{\displaystyle\int\limits_{\tau_{0}}^{\tau_{0}+\tau_{rev}}}
|\,\tau+\alpha\,\rangle\left(  \tau+\alpha\right)  \langle\,\tau
+\alpha\,|\,d\tau\\
&  =\exp\left(  -i\widehat{H}\alpha\right)  \left\{  \widehat{T}\left(
\tau_{0}\right)  +\alpha\widehat{I}\right\}  \exp\left(  i\widehat{H}%
\alpha\right)  ,
\end{align*}
where $\widehat{I}$ is the identity operation. But the same operator can be
expressed in another way:%
\[
\widehat{T}\left(  \tau_{0}+\alpha\right)  =\left\{
{\displaystyle\int\limits_{\tau_{0}+\alpha}^{\tau_{0}}}
+%
{\displaystyle\int\limits_{\tau_{0}}^{\tau_{0}+\tau_{rev}}}
+%
{\displaystyle\int\limits_{\tau_{0}+\tau_{rev}}^{\tau_{0}+\tau_{rev}+\alpha}}
\right\}  \,|\,\tau\,\rangle\tau\langle\,\tau\,|\,d\tau
\]
Exploiting the periodicity of the time states, we write the last integral on
the right as%
\[%
{\displaystyle\int\limits_{\tau_{0}+\tau_{rev}}^{\tau_{0}+\tau_{rev}+\alpha}}
|\,\tau\,\rangle\tau\langle\,\tau\,|\,d\tau=%
{\displaystyle\int\limits_{\tau_{0}}^{\tau_{0}+\alpha}}
|\,\tau\,\rangle\left(  \tau+\tau_{rev}\right)  \langle\,\tau\,|\,d\tau,
\]
leaving%
\[
\widehat{T}\left(  \tau_{0}+\alpha\right)  =\widehat{T}\left(  \tau
_{0}\right)  +\tau_{rev}%
{\displaystyle\int\limits_{\tau_{0}}^{\tau_{0}+\alpha}}
|\,\tau\,\rangle\langle\,\tau\,|\,d\tau
\]
Equating the two alternative forms for $\widehat{T}\left(  \tau_{0}%
+\alpha\right)  $ gives%
\begin{equation}
\exp\left(  -i\widehat{H}\alpha\right)  \widehat{T}\left(  \tau_{0}\right)
\exp\left(  i\widehat{H}\alpha\right)  +\alpha\widehat{I}=\widehat{T}\left(
\tau_{0}\right)  +\tau_{rev}%
{\displaystyle\int\limits_{\tau_{0}}^{\tau_{0}+\alpha}}
|\,\tau\,\rangle\langle\,\tau\,|\,d\tau\label{time operator (Heisenberg)}%
\end{equation}
Writing Eq.(\ref{time operator (Heisenberg)}) for infinitesimal $\alpha$ leads
to the commutation relation between $\widehat{T}\left(  \tau_{0}\right)  $ and
$\widehat{H}$:
\begin{equation}
\left[  \widehat{T}\left(  \tau_{0}\right)  ,\widehat{H}\right]
=i\widehat{I}-i\tau_{rev}|\,\tau_{0}\,\rangle\langle\,\tau_{0}\,|
\label{commutator-periodic}%
\end{equation}
\textit{Thus, while the time operator of Eq.(\ref{time operator}) exists for
any periodic system, it is never canonically conjugate to the Hamiltonian.}
Although a bit unsettling, this negative conclusion is an inevitable
consequence of periodicity, as has been argued persuasively by Pegg
\cite{Pegg}.

According to standard theory, the commutator of any operator with the
Hamiltonian dictates the time dependence of the associated observable. Applied
to the time operator $\widehat{T}$ and any normalized state $|\,\psi
(t)\,\rangle$, this principle combined with Eq.(\ref{commutator-periodic})
gives%
\begin{align}
\frac{d\tau_{avg}}{dt}  &  =-i\,\langle\,\psi(t)\,|\left[  \widehat{T}\left(
\tau_{0}\right)  ,\widehat{H}\right]  |\,\psi(t)\,\rangle\nonumber\\
&  =1-\tau_{rev}\left\vert \langle\,\tau_{0}\,|\,\psi(t)\,\rangle\right\vert
^{2}=1-\tau_{rev}\left\vert \langle\,\tau_{0}-t\,|\,\psi(0)\,\rangle
\right\vert ^{2} \label{system-time change}%
\end{align}
We see at once that average system time $\tau_{avg}$ faithfully tracks
laboratory time $t$ so long as the overlap $\langle\,\tau_{0}-t\,|\,\psi
(0)\,\rangle$ is negligible, but in periodic systems it is equally clear that
this happy state of affairs cannot last throughout an entire recurrence cycle.
Eq.(\ref{system-time change}) also implies that $\tau_{avg}$ is unchanging in
any stationary state, as expected for a time that is related to some [event] observable.

The discussion thus far leaves open the possibility that a canonical time
operator might still exist in aperiodic systems ($\tau_{rev}\rightarrow\infty
$), provided we choose $\tau_{0}$ judiciously. That hope is reinforced by the
observation that $\left\vert \langle\,\tau\,|\,\psi\,\rangle\right\vert ^{2}$
vanishes in the asymptotic regime for any normalizable state $|\,\psi
\,\rangle$ ($\langle\,\tau\,|\,\psi\,\rangle$ is square-integrable).
Explicitly, if $\tau_{0}$ scales with $\tau_{rev}$ and $\langle\,\tau
_{0}\,|\,\psi\,\rangle$ approaches zero `fast enough', the non-canonical term
\ in Eq.(\ref{commutator-periodic}) will vanish in the aperiodic limit.
Interestingly, the presence of even one truly discrete level in the spectrum
of $\widehat{H}$ precludes this possibility, since then
Eq.(\ref{history-mixed}) shows that $\tau_{rev}\left\vert \langle\,\tau
_{0}\,|\,\psi\,\rangle\right\vert ^{2}$ is $O(1)$ as $\tau_{rev}%
\rightarrow\infty$, no matter how we choose $\tau_{0}$. But \textit{absent any
isolated levels, the choice }$\tau_{0}\propto\tau_{rev}$ \textit{together with
the square-integrability of }$\langle\,\tau\,|\,\psi\,\rangle$\textit{ is
enough to ensure that the time operator of Eq.(\ref{time operator}) is
canonical to the Hamiltonian in the aperiodic limit }$\tau_{rev}%
\rightarrow\infty$. Having said this, however, we must caution that there is
no \textit{a-priori} guarantee that the integral of Eq.(\ref{time operator})
actually exists in this limit!

From an existence standpoint, the most forgiving choice for $\tau_{0}$ appears
to be\textit{ }$\tau_{0}=-\tau_{rev}/2$\textit{,} which leads to the
definition of a time operator in aperiodic systems as a Cauchy principal
value:%
\begin{equation}
\widehat{T}_{aperiodic}\equiv P%
{\displaystyle\int\limits_{-\infty}^{\infty}}
|\,\tau\,\rangle\tau\langle\,\tau\,|\,d\tau, \label{Cauchy-T}%
\end{equation}
The existence of $\widehat{T}_{aperiodic}$ hinges on the asymptotic behavior
of the timeline function $\langle\,\tau\,|\,\psi\,\rangle$. Generally, we can
expect asymptotic behavior consistent with the square-integrability of
$\langle\,\tau\,|\,\psi\,\rangle$ on $\left(  -\infty,\infty\right)  $, but
this alone is insufficient to secure the convergence of the integral in
Eq.(\ref{Cauchy-T}). As is well-known, the asymptotics of the (Fourier)
integral in Eq.(\ref{history-mixed}) are dictated by the properties of the
spectral wave function $\langle\,E\,|\,\psi\,\rangle$ \cite{Erdelyi}. We will
assume that $\langle\,E\,|\,\psi\,\rangle$ is continuously differentiable
throughout the continuum $E_{\min}\leq E\leq E_{\max}$, with a derivative that
is integrable over $E_{\min}<E<E_{\max}$. \textit{If,} \textit{additionally,
}$\langle\,E\,|\,\psi\,\rangle$\textit{ vanishes at the continuum edge(s), an
integration by parts of Eq.(\ref{history-mixed}) shows that }$\langle
\,\tau\,|\,\psi\,\rangle$\textit{ is }$o\left(  \tau^{-1}\right)  $\textit{,
and the integral of Eq.(\ref{Cauchy-T}) converges to define a valid time
operator for aperiodic systems.} As a corollary, we note that if the spectrum
has no natural bounds ($E_{\min}\rightarrow-\infty$ \textit{and} $E_{\max
}\rightarrow+\infty$), then the square-integrability of $\langle
\,E\,|\,\psi\,\rangle$ by itself is enough to guarantee a valid time operator.
Otherwise, the existence (or not) of $\widehat{T}_{aperiodic}$ involves more
delicate questions of convergence having to do with the Cauchy principle
value; such issues are best addressed in individual cases.

Finally, we examine how relevant properties of the spectral wave function
translate into the language of stationary state wave functions. To that end,
we write $\langle\,E\,|\,\psi\,\rangle$ in a generic coordinate basis whose
elements we denote simply as $|\,q\,\rangle$:%
\[
\langle\,E\,|\,\psi\,\rangle=%
{\displaystyle\int}
\psi(q)\,\langle\,q\,|\,E\,\rangle^{\ast}\,dq
\]
Here $\langle\,q\,|\,E\,\rangle$ is a stationary wave with energy $E$ and
$\psi(q)=\langle\,q\,|\,\psi\,\rangle$ is the Schr\"{o}dinger wave function
for the state $|\,\psi\,\rangle$. But this representation is valid for
\textit{any} normalizable state (the coordinate basis is complete), and so we
are free to choose $\psi(q)$ as we please, provided only that it is a
square-integrable function. Taking $\psi(q)$ to have support over only an
arbitrarily narrow interval about $q_{0}$ essentially `picks out' the
stationary wave value at $q_{0}$. In this way we argue that \textit{any demand
placed on }$\langle\,E\,|\,\psi\,\rangle$\textit{ for every [normalizable]
state }$|\,\psi\,\rangle$\textit{ becomes a condition on the
[energy-normalized] stationary wave }$\langle\,q\,|\,E\,\,\rangle$\textit{
that must be met for all }$\,q$.

\section{Example: Particle in Free Fall}

In this -- arguably the simplest -- case, we take $\widehat{H}=\widehat{p}%
^{2}/2m-F\widehat{x}$, with $F=-mg$ denoting the classical force of gravity.
(With $F=q\varepsilon$, the same Hamiltonian describes a charge $q$ in a
uniform electric field $\varepsilon$.) The spectrum is non-degenerate, and
stretches continuously from $E_{\min}=-\infty$ to $E_{\max}=\infty$. While the
unbounded nature of the spectrum from below is considered unphysical, this
model nonetheless serves a useful purpose by sidestepping the issue of
boundary conditions at the potential energy minimum. With no isolated levels
and an unbounded continuum, Eq.(\ref{history-mixed}) becomes
\begin{equation}
\langle\,\tau\,|\,\psi\,\rangle\equiv\frac{1}{\sqrt{2\pi}}%
{\displaystyle\int\limits_{-\infty}^{\infty}}
\exp\left(  iE\tau\right)  \langle\,E\,|\,\psi\,\rangle\,dE
\label{history-open}%
\end{equation}
Eq.(\ref{history-open}) will be recognized as a conventional Fourier
transform, the properties of which follow from the extensive theory on Fourier
integrals. In particular, the integral maps square-integrable functions
$\langle\,E\,|\,\psi\,\rangle$ into new functions $\langle\,\tau
\,|\,\psi\,\rangle$ (wave functions in the time basis) that are themselves
square-integrable \cite{Olver2}.

In the coordinate basis, the stationary states are $\langle x|\,E\,\rangle
=C_{E}Ai(-z)$, with $Ai(\ldots)$ the Airy function, $z\equiv\kappa\left(
x+E/F\right)  $, and $\kappa\equiv\left(  2mF\right)  ^{1/3}$ \cite{Landau}.
The constant $C_{E}$ is fixed such that these stationary states are
energy-normalized, i.e., $\langle\,E\,|\,E^{\prime}\,\rangle=\delta\left(
E\,-\,E^{\prime}\right)  $. From the orthogonality relation for Airy functions
\cite{Aspnes}, we find%
\begin{align*}
\langle\,E\,|\,E^{\prime}\,\rangle &  =C_{E}C_{E^{\prime}}%
{\displaystyle\int\limits_{-\infty}^{\infty}}
\,dx\,Ai(-\kappa x-\kappa E/F)\,Ai(-\kappa x-\kappa E^{\prime}/F)\\
&  =C_{E}^{2}\frac{1}{\kappa}\delta\left(  \kappa E/F\,-\kappa E^{\prime
}/F\right)  =C_{E}^{2}\frac{\left\vert F\right\vert }{\kappa^{2}}\delta\left(
E\,-E^{\prime}\right)
\end{align*}
Thus, the desired normalization follows if we take%
\begin{equation}
C_{E}=\sqrt{\kappa^{2}/\left\vert F\right\vert } \label{norm-freefall}%
\end{equation}

\subsection{Free-Fall Timelines}

The Schr\"{o}dinger wave function associated with a time state follows by
taking $|\,\psi\,\rangle=|\,x\,\rangle$ in Eq.(\ref{history-open}). Defining
$\langle\,x\,|\,\tau\,\rangle\equiv\Xi_{\tau}(x)$ , we find with the help of
Eq.(\ref{norm-freefall})%
\begin{equation}
\Xi_{\tau}(x)=\sqrt{\frac{\kappa^{2}}{2\pi\left\vert F\right\vert }}%
{\displaystyle\int\limits_{-\infty}^{\infty}}
\,dE\exp\left(  -iE\tau\right)  \,Ai(-\kappa x-\kappa E/F)
\label{integral-freefall}%
\end{equation}
The integral is essentially the Fourier transform of the Airy function; this
is readily identified from the integral representation for $Ai$
\cite{Abramowitz} to give%
\begin{equation}
\Xi_{\tau}(x)=\sqrt{\frac{\left\vert F\right\vert }{2\pi}}\exp\left(
iFx\tau-iF^{2}\tau^{3}/6m\right)  \label{waves-open}%
\end{equation}
Notice that $\Xi_{\tau}^{\ast}(x)=\Xi_{-\tau}(x)$, a result that also follows
from inspection of the integral form, Eq.(\ref{integral-freefall}). The
timeline wave for this case is simplicity itself: except for a [physically
insignificant] phase factor and a different normalization,
Eq.(\ref{waves-open}) is the usual plane wave associated with the momentum
eigenstate $|\,p\,\rangle$ for momentum $p=F\tau$! Accordingly, at the system
time $\tau$ the particle attains a specified value of momentum ($F\tau$).

Routine -- albeit not rigorous -- means for establishing directly the
properties of the resulting timeline rely on the integral representation of
the Dirac delta function%
\begin{equation}
\delta\left(  k\right)  =\frac{1}{2\pi}%
{\displaystyle\int\limits_{-\infty}^{\infty}}
\exp\left(  ikx\right)  \,\,dx \label{Dirac delta}%
\end{equation}
coupled with a certain flair for manipulation. For example, using
Eq.(\ref{Dirac delta}) we easily discover that the timeline waves for this
case are truly orthogonal:%
\begin{align*}
\langle\,\tau^{\prime}\,|\,\tau\,\rangle &  \equiv%
{\displaystyle\int\limits_{-\infty}^{\infty}}
\,dx\,\Xi_{\tau^{\prime}}^{\ast}(x)\,\Xi_{\tau}(x)\\
&  =\frac{\left\vert F\right\vert }{2\pi}\exp\left[  iF^{2}\left(
\tau^{\prime3}-\tau^{3}\right)  /6m\right]
{\displaystyle\int\limits_{-\infty}^{\infty}}
\,dx\,\exp\left[  iFx\left(  \tau-\tau^{\prime}\right)  \right]  =\delta
(\tau^{\prime}-\tau)
\end{align*}
Timeline closure (cf. Eq.(\ref{closure rule})) can be confirmed with equal
ease:%
\[%
{\displaystyle\int\limits_{-\infty}^{\infty}}
\,d\tau\,\Xi_{\tau}^{\ast}(x^{\prime})\,\Xi_{\tau}(x)=\frac{\left\vert
F\right\vert }{2\pi}%
{\displaystyle\int\limits_{-\infty}^{\infty}}
\,d\tau\,\exp\left[  iF\tau\left(  x-x^{\prime}\right)  \right]
=\delta(x-x^{\prime})
\]
These cavalier manipulations find their ultimate justification in the theory
of distributions, or generalized functions, which gives precise meaning to
integrals such as Eq.(\ref{Dirac delta}) that do not converge in any standard sense.

\subsection{Time Operator for a Freely-Falling Particle}

With a spectrum that is unbounded both above and below, a particle in
free-fall is described by timeline functions $\langle\,\tau\,|\,\psi\,\rangle$
that support a canonical time operator, as discussed in Sec. III. Indeed,
Eq.(\ref{history-open}) can be integrated once by parts to show that
$\langle\,\tau\,|\,\psi\,\rangle$ is $o\left(  \tau^{-1}\right)  $ as
$\tau\rightarrow\pm\infty$, just enough to secure convergence of the integral
in Eq.(\ref{Cauchy-T}). The restrictions leading to this conclusion are quite
modest ($\langle\,E\,|\,\psi\,\rangle$\ must be continuously differentiable
and its derivative integrable over the entire real line), and likely to be met
in all but the most pathological cases. Notice that the existence of a time
operator here does \textit{not} contradict Pauli's argument, since all
energies are allowed for a freely-falling mass.

With still more manipulative flair, we can proceed to assign matrix elements
of $\widehat{T}$ in the coordinate basis (cf. Eq.(\ref{time operator})):%
\begin{align*}
\langle\,x\,|\widehat{T}|\,x^{\prime}\rangle &  =%
{\displaystyle\int\limits_{-\infty}^{\infty}}
\langle\,x\,|\,\tau\,\rangle\tau\langle\,\tau\,|\,x^{\prime}\rangle\,d\tau=%
{\displaystyle\int\limits_{-\infty}^{\infty}}
\,d\tau\,\Xi_{\tau}^{\ast}(x^{\prime})\tau\,\Xi_{\tau}(x)\\
&  =\frac{\left\vert F\right\vert }{2\pi}%
{\displaystyle\int\limits_{-\infty}^{\infty}}
\,d\tau\,\tau\exp\left[  iF\tau\left(  x-x^{\prime}\right)  \right]  =\frac
{1}{iF}\frac{\partial}{\partial x}\delta(x-x^{\prime})
\end{align*}
These are reminiscent of matrix elements of the momentum operator
$\widehat{p}$ in this basis: comparing the two, we arrive at the
identification%
\begin{equation}
\widehat{T}_{free-fall}=\frac{1}{F}\widehat{\,p} \label{T-freefall}%
\end{equation}
By inspection, the time operator of Eq.(\ref{T-freefall}) clearly is Hermitian
and canonically conjugate to $\widehat{H}$, $\left[  \widehat{T}%
,\widehat{H}\right]  =i$. Thus,\textit{ the canonical time operator for a
freely-falling particle is simply a scaled version of the operator for
particle momentum!}

\section{Example: Free Particle in One Dimension}

The Hamiltonian for this case $\widehat{H}=\widehat{p}^{2}/2m$ describes a
particle free to move along the line $-\infty<x<\infty$. The spectrum of
$\widehat{H}$\ extends from $E_{\min}=0$ to $E_{\max}=\infty$, and each energy
level is doubly-degenerate. Accordingly, the timeline waves in this example
$\langle\,\tau\,|\,\psi\,\rangle$\ are calculated from the expression (cf.
Eq.(\ref{history-mixed}))%
\begin{equation}
\langle\,\tau\,|\,\psi\,\rangle\equiv\frac{1}{\sqrt{2\pi}}%
{\displaystyle\int\limits_{0}^{\infty}}
\exp\left(  iE\tau\right)  \langle\,E\,|\,\psi\,\rangle\,dE
\label{history-semiopen}%
\end{equation}
(The two-fold degeneracy of the free-particle continuum implies that the
stationary states carry an additional label, as elaborated below.)
Eq.(\ref{history-semiopen}) is a \textit{holomorphic Fourier transform}, with
very close ties to the standard Fourier transform encountered in Section IV.
Indeed, if we agree to extend the function $\langle\,E\,|\,\psi\,\rangle$ to
all real energies by the rule $\langle\,E\,|\,\psi\,\rangle\equiv0$ for $E<0$,
then Eq.(\ref{history-semiopen}) reverts to the familiar Fourier integral. For
square-integrable functions $\langle\,E\,|\,\psi\,\rangle$, Fourier integral
theory then guarantees that the transform function $\langle\,\tau
\,|\,\psi\,\rangle$ also is square-integrable over its domain, $-\infty
<\tau<\infty$ \cite{Olver2}. Furthermore, the inverse transform is%
\begin{equation}
\langle\,E\,|\,\psi\,\rangle=\frac{1}{\sqrt{2\pi}}%
{\displaystyle\int\limits_{-\infty}^{\infty}}
\exp\left(  -iE\tau\right)  \langle\,\tau\,|\,\psi\,\rangle\,d\tau
\label{spectrum-semiopen}%
\end{equation}
The essential new feature introduced by a spectrum bounded from below is that
$\langle\,\tau\,|\,\psi\,\rangle$ calculated from Eq.(\ref{history-semiopen})
is analytic (holomorphic) for all complex values of $\tau$ in the upper
half-plane $\Im m\,\tau>0$, and vanishes as $\left\vert \tau\right\vert
\rightarrow\infty$ in the entire sector $0\leq\arg(\tau)\leq\pi$. In turn,
these properties of $\langle\,\tau\,|\,\psi\,\rangle$ in the complex plane
ensure that $\langle\,E\,|\,\psi\,\rangle$ calculated from
Eq.(\ref{spectrum-semiopen}) is truly zero for all negative values of $E$
(follows from applying the residue calculus to a contour of integration
consisting of the real axis closed by an infinite semicircle in the upper half-plane).

\subsection{Free-Particle Timelines}

We begin by taking the degenerate eigenfunctions to be plane waves, writing
$|\,E\,\rangle\rightarrow|\,k\,\rangle$ with $\langle\,x\,|\,k\,\rangle
=C_{k}\exp\left(  ikx\right)  $. These are harmonic oscillations with
wavenumber $k$ and energy $E_{k}=k^{2}/2m$. Orthogonality of these waves is
expressed by%
\begin{align*}
\langle\,k\,|\,k^{\prime}\,\rangle &  =C_{k}C_{k^{\prime}}%
{\displaystyle\int\limits_{-\infty}^{\infty}}
\,dx\,\exp\left(  ikx-ik^{\prime}x\right) \\
&  =C_{k}^{2}2\pi\,\delta\left(  k-k^{\prime}\right)  =C_{k}^{2}\,\frac
{2\pi\left\vert k\right\vert }{m}\,\delta\left(  E_{k}\,-\,E_{k^{\prime}%
}\right)  ,
\end{align*}
so that energy normalization in this case requires%
\begin{equation}
C_{k}=\sqrt{\frac{m}{2\pi\left\vert k\right\vert }} \label{norm-free}%
\end{equation}

Plane waves running in opposite directions ($\pm k$) give rise to distinct
quantum histories, which we distinguish by the direction of wave propagation:
$|\,\tau\,\rangle\rightarrow|\,\tau,\rightleftarrows\rangle$. Timeline
elements in this representation are described by the Schr\"{o}dinger wave
functions $\langle\,x\,|\,\tau,\rightleftarrows\rangle\equiv\Xi_{\tau
}^{\rightleftarrows}(x)$, obtained by taking $|\,\psi\,\rangle=|\,x\,\rangle$
in Eq.(\ref{spectrum-semiopen}):%
\begin{align}
\Xi_{\tau}^{\rightleftarrows}(x)  &  =\frac{1}{\sqrt{2\pi}}%
{\displaystyle\int\limits_{0}^{\infty}}
\exp\left(  -iE_{k}\tau\right)  C_{k}\exp(\pm ikx)\,dE_{k}\nonumber\\
&  =\frac{1}{2\pi\sqrt{m}}%
{\displaystyle\int\limits_{0}^{\infty}}
\,dk\,\sqrt{k}\exp\left(  \pm ikx-ik^{2}\tau/2m\right)
\label{integral-directional}%
\end{align}
In this and subsequent expressions, the right (left) arrow is associated with
the upper (lower) sign. The integral of Eq.(\ref{integral-directional}) is
related to the parabolic cylinder function $D_{\nu}(\ldots)$; in particular,
we have for $\Im m\,\tau<0$ ($\Re e\left(  \,i\tau\right)  >0$)
\cite{Gradshteyn2}%
\begin{equation}
\Xi_{\tau}^{\rightleftarrows}(x)=\frac{1}{4\sqrt{\pi m}}z^{3/2}\exp\left(
-x^{2}z^{2}/4\right)  D_{-3/2}\left(  \mp ixz\right)  \text{\qquad}%
z\equiv\sqrt{\frac{m}{i\tau}} \label{waves-directional}%
\end{equation}
This form holds for $\left\vert \arg\,z\right\vert <\pi/4$, but since $D_{\nu
}$ is an entire function of its argument \cite{Bateman2} the result can be
analytically continued to all real values of $\tau$ ($\arg\,z=\pm\pi/4$).
Clearly, $\Xi_{\tau}^{\rightleftarrows}(-x)=\Xi_{\tau}^{\leftrightarrows}(x)$.
For $\tau$ real and negative we take $\arg\,z=\pi/4$ in
Eq.(\ref{waves-directional}) to obtain%
\begin{align}
\Xi_{-\tau}^{\rightleftarrows}(x)  &  =\frac{1}{4\sqrt{\pi m}}\left\vert
z\right\vert ^{3/2}\exp\left(  -ix^{2}\left\vert z\right\vert ^{2}%
/4+i3\pi/8\right)  D_{-3/2}\left(  \mp ix\left\vert z\right\vert \exp\left(
+i\pi/4\right)  \right) \nonumber\\
&  =\left[  \Xi_{\tau}^{\leftrightarrows}(x)\right]  ^{\ast},
\label{time-symmetry}%
\end{align}
a relation that also is evident from the integral representation,
Eq.(\ref{integral-directional}). These results are consistent with the
pioneering 1974 work of Kijowski \cite{Kijowski}, who used an axiomatic
approach to construct a distribution of arrival times in the momentum
representation; however, the coordinate form $\Xi_{\tau}^{\rightleftarrows
}(x)$ given by Eq.(\ref{waves-directional}) did not appear in the literature
until more than twenty years later \cite{Muga}.

Another representation better suited to numerical computation relies on the
degeneracy of free-particle waves to construct histories from standing wave
combinations of plane waves. Since standing waves are parity eigenfunctions,
parity -- not direction of travel -- is the `good' quantum label in this
scheme. The competing descriptions in terms of running waves and standing
waves are connected by a unitary transformation; as noted in Sec. II, this
same transformation also relates the timelines stemming from the two
representations (cf. Eq.(\ref{subspace-rotation})):%
\begin{equation}%
\begin{bmatrix}
|\,\tau,\rightarrow\rangle\\
|\,\tau,\leftarrow\rangle
\end{bmatrix}
=\frac{1}{\sqrt{2}}%
\begin{bmatrix}
1 & i\\
1 & -i
\end{bmatrix}%
\begin{bmatrix}
|\,\tau,+\,\rangle\\
|\,\tau,-\,\rangle
\end{bmatrix}
\label{dir-parity map}%
\end{equation}
As it happens, standing waves are simply related to Bessel functions
$J_{\alpha}$ of order $\alpha=\pm1/2$. Using $\exp(\pm ikx)=\sqrt{\pi
kx/2}\left[  J_{-1/2}(kx)\pm iJ_{\mp1/2}(kx)\right]  $ in
Eq.(\ref{integral-directional}), we find on comparing with
Eq.(\ref{dir-parity map}) that timeline elements in the standing-wave
representation are described by the coordinate-space forms $\langle
\,x\,|\,\tau,\pm\rangle\equiv\Xi_{\tau}^{\pm}(x)$, where%
\begin{equation}
\Xi_{\tau}^{\pm}(x)=\frac{\sqrt{x}}{2\sqrt{\pi m}}%
{\displaystyle\int\limits_{0}^{\infty}}
\,dk\,k\exp\left(  -ik^{2}\tau/2m\right)  J_{\mp1/2}(kx)\text{\qquad}x\geq0
\label{integral-parity}%
\end{equation}
The sign label $\left(  \pm\right)  $ specifies the parity of these waves and
prescribes their extension to $x<0$.

Once again the integrals in Eq.(\ref{integral-parity}) can be evaluated in
closed form. The odd-parity timeline waves for $\tau>0$ and $x\geq0$ are given
by \cite{Gradshteyn}
\begin{equation}
\Xi_{\tau}^{-}(x)=\sqrt{\frac{2}{m}}\left(  xz\right)  ^{3/2}\exp\left(
ix^{2}z-i\pi/8\right)  \,\left[  J_{3/4}\left(  x^{2}z\right)  -iJ_{-1/4}%
\left(  x^{2}z\right)  \right]  \text{\qquad}z\equiv\frac{m}{4\tau}
\label{waves-free-o}%
\end{equation}
Unlike Eq.(\ref{waves-directional}), $z$ in this expression is real and
positive. Eq.(\ref{waves-free-o}) is essentially the result reported in a
recent paper by Galapon et. al. \cite{Galapon}.

The odd-parity states by themselves constitute a complete history for an
otherwise free particle that is confined to the half-axis $x>0$ (e.g., by an
infinite potential wall at the origin), but for a truly free particle we also
need the even-parity states. The even-parity timeline waves for $\tau>0$ and
$x\geq0$ are \cite{Gradshteyn}
\begin{equation}
\Xi_{\tau}^{+}(x)=\sqrt{\frac{2}{m}}\left(  xz\right)  ^{3/2}\exp\left(
ix^{2}z+i\pi/8\right)  \,\left[  J_{1/4}\left(  x^{2}z\right)  -iJ_{-3/4}%
\left(  x^{2}z\right)  \right]  \text{\qquad}z\equiv\frac{m}{4\tau}
\label{waves-free-e}%
\end{equation}
Eqs.(\ref{waves-free-o}) and (\ref{waves-free-e}) are valid for $\tau>0$;
results for $\tau<0$ follow from $\Xi_{-\tau}^{\pm}(x)=\left[  \Xi_{\tau}%
^{\pm}(x)\right]  ^{\ast}$ (cf. Eq.(\ref{integral-parity})). Timeline waves of
either parity are well-behaved for all finite values of $x$, but diverge (as
$\left\vert x\right\vert ^{1/2}$) for $\left\vert x\right\vert \rightarrow
\infty$.

The time states constructed from running waves admit an interesting physical
interpretation. For any $\tau>0$ the rightward-running timeline wave
$\Xi_{\tau}^{\rightarrow}(x)$ diverges as $x^{1/2}$ for $x\rightarrow\infty$,
but vanishes as $\left\vert x\right\vert ^{-3/2}$ for $x\rightarrow-\infty$;
more precisely, the asymptotics of the parabolic cylinder function
\cite{Bateman3} show that for any $\tau>0$ ($\arg\,z=-\pi/4$) and large
$\left\vert x\right\vert $%
\[
\Xi_{\tau}^{\rightarrow}(x)\sim%
\begin{array}
[c]{cc}%
\exp\left(  imx^{2}/2\tau\right)  x^{1/2} & \qquad x>0\\
\left\vert x\right\vert ^{-3/2} & \qquad x<0
\end{array}
\]
These features are reversed for the leftward-running wave $\Xi_{\tau
}^{\leftarrow}(x)$. (Corresponding results for $\tau<0$ follow directly from
the relation $\Xi_{-\tau}^{\rightleftarrows}(x)=\left[  \Xi_{\tau
}^{\leftrightarrows}(x)\right]  ^{\ast}$.) The changeover in behavior occurs
in a small neighborhood of $x=0$. The width of this \textit{transition region}
narrows with diminishing values of $\tau$, approaching zero for $\tau=0$. The
Schr\"{o}dinger waveform $\Xi_{\tau}^{\rightarrow}(x)$ for two [system] times
straddling $\tau=0$\ are shown as Figs. 1 and 2, using a color-for-phase
plotting style that captures both the modulus and phase of these
complex-valued functions. The functions at any [laboratory] time other than
$t=0$ are found by replacing $|\,\tau\,\rangle$ in the above argument with the
evolved state $\exp\left(  -i\widehat{H}t\right)  |\,\tau\,\rangle
=|\,\tau\,+t\,\rangle$. Thus, for $t\neq0$ the function behavior shown in the
figures is unaltered, but the origin of time is shifted so that the abrupt
change in behavior around $x=0$ occurs generally at the [system] time
$\tau=-t$. Consequently, $-\tau$ is designated an \textit{arrival time},
inasmuch as it signals the [laboratory] time when the bulk of probability
shifts from one side of the coordinate origin to the other \cite{Muga}. (The
minus sign can be understood by noting that as system time increases, the time
to arrival diminishes.) In summary, the construction of
Eq.(\ref{waves-directional}) leads in this case to \textit{time-of-arrival
states} for leftward [$\leftarrow$] or rightward [$\rightarrow$]-running
waves, with $-\tau$ specifying the arrival time at the coordinate origin
$x=0$. This interpretation receives further support from the recent work of
Galapon \cite{Galapon}, who showed that similar states in a confined space
(where they can be normalized) are such that the event of the centroid
arriving at the origin coincides with the uncertainty in position being
minimal. Arguably this is the best we can do in defining arrival times for
entities subject to quantum uncertainty.

Time-of-arrival states specific to an arbitrary coodinate point, say $x=a$,
can be obtained as spatial translates of those constructed here:
$|\,\tau,\rightleftarrows,a\,\rangle=\exp\left(  -i\widehat{p}a\right)
|\,\tau,\rightleftarrows\,\rangle$ ($\widehat{p}$, the particle momentum
operator, is the generator of displacements). The associated Schr\"{o}dinger
wave function is $\langle\,x\,|\,\tau,\rightleftarrows,a\rangle=\langle
\,x-a\,|\,\tau,\rightleftarrows\rangle=\Xi_{\tau}^{\rightleftarrows}(x-a)$. In
keeping with our earlier observation concerning the phase ambiguity of
timelines, we note that spatial translates also can be recovered from
Eq.(\ref{integral-directional}) by re-defining the phases of the stationary
waves as $\langle\,x\,|\,k\,\rangle\rightarrow\exp\left(  \mp ika\right)
\langle\,x\,|\,k\,\rangle$.
\begin{figure}[ptb]%
\centering
\includegraphics[
natheight=3.957400in,
natwidth=5.389500in,
height=2.4128in,
width=3.403in
]%
{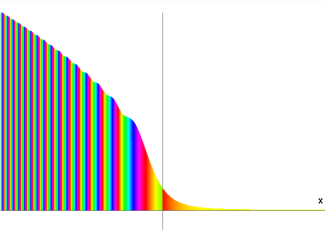}%
\caption{The timeline waveform $\Xi_{\tau}^{\rightarrow}(x)$ for $\tau
=-0.005$\ constructed from rightward-running plane waves. The shading
(coloring) represents varying phase values for this complex function. In units
where $\hbar=m=1$, the plot extends from $x=-1$ to $x=+1$. Except for an
overall phase factor, this also represents the conjugate of the waveform
$\Xi_{\tau}^{\leftarrow}(x)$ at the system time $\tau=0.005$ (cf.
Eq.(\ref{time-symmetry})).}%
\label{fig1}%
\end{figure}
\begin{figure}[ptb]%
\centering
\includegraphics[
natheight=3.957400in,
natwidth=5.389500in,
height=2.4025in,
width=3.403in
]%
{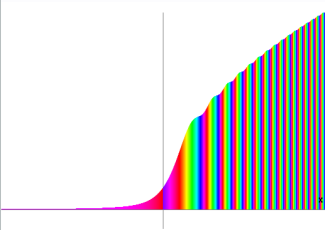}%
\caption{The timeline waveform $\Xi_{\tau}^{\rightarrow}(x)$ for $\tau
=0.005$\ constructed from rightward-running plane waves. The shading
(coloring) represents varying phase values for this complex function. In units
where $\hbar=m=1$, the plot extends from $x=-1$ to $x=+1$. Except for an
overall phase factor, this also represents the conjugate of the waveform
$\Xi_{\tau}^{\leftarrow}(x)$ at the system time $\tau=-0.005$ (cf.
Eq.(\ref{time-symmetry})).}%
\label{fig2}%
\end{figure}
\bigskip

\subsection{A Free-Particle Time Operator in One Dimension}

A time operator for free particles can be constructed following the recipe of
Sec. III. The invariance expressed by Eq.(\ref{subspace-projector}) ensures
that the same time operator results no matter which [degenerate]
representation we choose for the computation. With parity as the `good'
quantum number, the free-particle time operator is composed from operators in
the even- and odd-parity subspaces: $\widehat{T}=\widehat{T}_{+}%
\oplus\widehat{T}_{-}$, where (cf. Eq.(\ref{Cauchy-T}))%
\begin{equation}
\widehat{T}_{\pm}\equiv P%
{\displaystyle\int\limits_{-\infty}^{\infty}}
|\,\tau,\pm\rangle\tau\langle\,\tau,\pm|\,d\tau
\end{equation}
Now the energy-normalized stationary waves of odd parity vanish at the lower
spectral bound as $E^{1/4}$(cf. Eq.(\ref{norm-free})), so the general theory
of Sec. III implies that $\widehat{T}_{-}$ is well-defined by the integral
above, the principal value notwithstanding. The coordinate-space matrix
elements of this operator are simply related to one of a class of integrals
$I_{l}\left(  r,r^{\prime}\right)  $ studied in Appendix A; using the result
reported there, we find
\begin{equation}
\langle\,x\,|\widehat{T}_{-}|\,x^{\prime}\rangle=%
{\displaystyle\int\limits_{-\infty}^{\infty}}
\left[  \Xi_{\tau}^{-}(x)\right]  ^{\ast}\tau\Xi_{\tau}^{-}(x^{\prime}%
)\,d\tau=\frac{1}{2}xx^{\prime}I_{0}\left(  x,x^{\prime}\right)  =i\frac{m}%
{4}x_{<}\operatorname{sgn}\left(  x-x^{\prime}\right)
\end{equation}

The case for $\widehat{T}_{+}$ is more delicate, since the energy-normalized
stationary waves of even parity actually diverge at the lower spectral bound
as $E^{-1/4}$ (cf. Eq.(\ref{norm-free})). Nonetheless, Appendix A confirms
that the principal value integral for the coordinate space matrix elements of
$\widehat{T}_{+}$ remains well-defined, and can be evaluated in closed form to
give%
\begin{equation}
\langle\,x\,|\widehat{T}_{+}|\,x^{\prime}\rangle=%
{\displaystyle\int\limits_{-\infty}^{\infty}}
\left[  \Xi_{\tau}^{+}(x)\right]  ^{\ast}\tau\Xi_{\tau}^{+}(x^{\prime}%
)\,d\tau=\frac{1}{2}xx^{\prime}I_{-1}\left(  x,x^{\prime}\right)  =i\frac
{m}{4}x_{>}\operatorname{sgn}\left(  x-x^{\prime}\right)
\end{equation}
Combining the even and odd-parity computations, we arrive at the provocatively
simple form%
\begin{equation}
\langle\,x\,|\widehat{T}_{1d-free}|\,x^{\prime}\rangle=i\frac{m}{4}\left(
x+x^{\prime}\right)  \operatorname{sgn}\left(  x-x^{\prime}\right)
\label{T-1d-free}%
\end{equation}
Eq.(\ref{T-1d-free}) agrees with the formula reported by Galapon et. al.
\cite{Galapon} for a particle confined to a section of the real line, in the
limit where the domain size becomes infinite. Here we arrive at the same
result in an unbounded space using an alternative limiting process -- the
accessible states model.

\section{Example: Free Particle in Three Dimensions}

In this case, the Hamiltonian $\widehat{H}$ is the operator for kinetic energy
in a three-dimensional space. The spectrum of $\widehat{H}$ is semi-infinite
(bounded from below by $E=0$, but no upper limit) and composed of degenerate
levels. This degeneracy breeds multiple timelines, conveniently indexed by the
same quantum numbers that label the spectral states. Again there is some
flexibility in labeling here depending upon what dynamical variables we opt to
conserve along with particle energy, but the general timeline wave
$\langle\,\tau\,|\,\psi\,\rangle$ is constructed from its spectral counterpart
$\langle\,E\,|\,\psi\,\rangle$ following the same prescription used in the
one-dimensional case, Eq.(\ref{history-semiopen}).

\subsection{Angular Momentum Timelines for a Free Particle}

In the angular momentum representation, the stationary states are indexed by a
continuous wave number $k$ (any non-negative value), an orbital quantum number
$l$ (a non-negative integer), and a magnetic quantum number $m_{l}$ (an
integer between $-l$ and $+l$, not to be confused with particle mass):
$|\,E\,\rangle\rightarrow|\,klm_{l}\,\rangle$. This stationary state has
energy $E_{k}=k^{2}/2m$. The associated Schr\"{o}dinger waveforms are
spherical waves $\langle\overrightarrow{r}\,|\,klm_{l}\,\rangle=$
$C_{k}\,\,j_{l}(kr)Y_{l}^{m_{l}}(\Omega_{r})$, formed as a product of a
spherical Bessel function $j_{l}$ with a spherical harmonic $Y_{l}^{m_{l}}$.
$C_{k}$ is a constant that -- for the construction of timelines -- is fixed by
energy normalization. Noting that the spherical harmonics are themselves
normalized over the unit sphere, we apply the Bessel function closure rule
\cite{Arfken} to evaluate the remaining portion of the normalization integral:%
\begin{align*}
\langle\,klm_{l}\,|\,k^{\prime}lm_{l}\rangle &  =C_{k}C_{k^{\prime}}%
{\displaystyle\int\limits_{0}^{\infty}}
\,dr\,r^{2}j_{l}\left(  kr\right)  j_{l}\left(  k^{\prime}r\right) \\
&  =C_{k}^{2}\frac{\pi}{2k^{2}}\,\delta\left(  k-k^{\prime}\right)  =C_{k}%
^{2}\,\frac{\pi}{2mk}\,\delta\left(  E_{k}\,-\,E_{k}^{\prime}\right)
\end{align*}
Thus, energy normalization of these spherical waves requires%
\begin{equation}
C_{k}=\sqrt{\frac{2mk}{\pi}} \label{norm-spherical}%
\end{equation}

The time states $|\,\tau lm_{l}\rangle$ in this representation have components
in the coordinate basis given by $\langle\overrightarrow{r}\,|\,\tau
lm_{l}\,\rangle\equiv\Xi_{\tau}^{l}(r)Y_{l}^{m_{l}}(\Omega_{r})$ where
$\Xi_{\tau}^{l}(r)$, the radial piece of the timeline wave, is calculated from
(cf. Eq.(\ref{history-semiopen})):%
\begin{align}
\Xi_{\tau}^{l}(r)  &  =\frac{1}{\sqrt{2\pi}}%
{\displaystyle\int\limits_{0}^{\infty}}
\exp\left(  -iE_{k}\tau\right)  \,C_{k}\,j_{l}\left(  kr\right)
\,dE_{k}\nonumber\\
&  =\frac{1}{\sqrt{2\pi mr}}%
{\displaystyle\int\limits_{0}^{\infty}}
\exp\left(  -ik^{2}\tau/2m\right)  \,kJ_{l+1/2}\left(  kr\right)  \,dk
\label{integral-spherical}%
\end{align}
The last line follows from the connection between spherical Bessel functions
and the (cylinder) Bessel functions of the first kind. The closure rule obeyed
by these time states%
\[%
{\displaystyle\int\limits_{-_{\infty}}^{\infty}}
\,\Xi_{\tau}^{l}(r)\,\left[  \Xi_{\tau}^{l}(r^{\prime})\right]  ^{\ast}%
d\tau=\frac{1}{r^{2}}\,\delta\left(  r^{\prime}-r\right)
\]
can be confirmed from the integral representation of
Eq.(\ref{integral-spherical}) using the closure rule for Bessel functions
\cite{Arfken}.

The integral in Eq.(\ref{integral-spherical}) converges for all $\Im
m\,\tau\leq0$ and any $l\geq0$. Defining $2\alpha\equiv l-1/2$, we find
\cite{Gradshteyn}%
\begin{equation}
\Xi_{\tau}^{l}(r)=\sqrt{\frac{4r}{m}}z^{3/2}\exp\left(  ir^{2}z-i\pi
\frac{2\alpha+1}{4}\right)  \,\left[  J_{\alpha+1}\left(  r^{2}z\right)
-iJ_{\alpha}\left(  r^{2}z\right)  \right]  \text{\qquad}z\equiv\frac{m}%
{4\tau} \label{waves-BesselJ}%
\end{equation}
For fixed $\alpha$, $J_{\alpha}(\ldots)$ is a regular function of its argument
throughout the complex plane cut along the negative real axis. Thus, through
the magic of analytic continuation, Eq.(\ref{waves-BesselJ}) extends
$\Xi_{\tau}^{l}(r)$ to the whole cut $z$-plane$~\left\vert \arg(z)\right\vert
<\pi$. Now for any real $\tau>0$, $z$ is a positive number, say $z=x$. To
recover results for $\tau<0$, $z$ must approach the negative real axis from
\textit{above} ($\arg(z)\rightarrow\pi$ for $\arg(\tau)\rightarrow-\pi$).
Writing $z=x\exp\left(  i\pi\right)  $ in Eq.(\ref{waves-BesselJ}) and using
$J_{\alpha}(\exp\left(  i\pi\right)  x)=\exp\left(  i\pi\alpha\right)
J_{\alpha}(x)$ \cite{Bateman} leads to the relation
\begin{equation}
\Xi_{-\tau}^{l}(r)=\left[  \Xi_{\tau}^{l}(r)\right]  ^{\ast}%
\end{equation}
for any real value of $\tau$, a result that also is evident from the integral
form, Eq.(\ref{integral-spherical}).

The behavior of $\Xi_{\tau}^{l}(r)$ for small $r$ and/or large $\tau$ follows
directly from the power series representation of the Bessel function
\cite{Abramowitz2}. Apart from numerical factors, we find from
Eq.(\ref{waves-BesselJ})%
\begin{equation}
\Xi_{\tau}^{l}(r)\simeq\frac{z^{3/2+\alpha}}{r^{5/2}}\,\sim\frac{r^{l}}%
{\tau^{l/2+5/4}}\text{\qquad}r^{2}\left\vert z\right\vert \ll1
\end{equation}
and this result is valid in any sector of the cut $z$-plane. Similarly, the
asymptotic series for the Bessel function \cite{Abramowitz3} furnishes a
large-argument approximation to $\Xi_{\tau}^{l}(r)$, valid for any $l\geq0$
and $\left\vert \arg(z)\right\vert <\pi$:
\begin{equation}
\Xi_{\tau}^{l}(r)\sim\frac{1}{r^{5/2}}\sqrt{\frac{2}{\pi m}}\left[  \left(
\frac{l\left(  l+1\right)  +1/4}{4}-i2z\right)  \exp\left(  i2z-i\pi
\frac{2l+1}{4}\right)  +\frac{2l+1}{4}\right]  \text{\qquad}r^{2}\left\vert
z\right\vert \gg1
\end{equation}

\subsection{Uni-Directional Timelines for a Free Particle}

Free particles also can be described by momentum eigenstates labeled by a wave
vector $\overrightarrow{k}$. These momentum states have energy\ $E_{k}%
=k^{2}/2m$, and so must be expressible as a superposition of angular momentum
states with the same energy:%
\begin{equation}
|\,\overrightarrow{k}\,\rangle=%
{\displaystyle\sum_{l=0}^{\infty}}
{\displaystyle\sum_{\,m_{l}=-l}^{l}}
U_{l}^{m_{l}}\left(  \widehat{k}\right)  |\,klm_{l}\,\rangle
\label{sph-2-plane transform}%
\end{equation}
Here $\widehat{k}$\ is the unit vector specifying the orientation of the wave
vector with modulus $k$. The transformation from the angular momentum
representation to the linear one should be unitary to preserve the energy
normalization required for the construction of timelines. To identify the
transformation coefficients $U_{l}^{m_{l}}\left(  \widehat{k}\right)  $, we
note first that the Schr\"{o}dinger waveforms associated with
$|\,\overrightarrow{k}\,\rangle$\ are plane waves multiplied by a suitable
normalizing constant $C_{k}^{(uni)}$:%
\begin{equation}
\langle\overrightarrow{r}\,|\,\overrightarrow{k}\,\rangle=C_{k}^{(uni)}%
\,\,\exp\left(  i\overrightarrow{k}\cdot\overrightarrow{r}\right)
\label{plw-free3d}%
\end{equation}
Next, we appeal to the spherical wave decomposition of a plane wave
\cite{Newton}%
\[
\exp\left(  i\overrightarrow{k}\cdot\overrightarrow{r}\right)  =4\pi%
{\displaystyle\sum_{l=0}^{\infty}}
i^{l}j_{l}(kr)%
{\displaystyle\sum_{\,m_{l}=-l}^{l}}
Y_{l}^{m_{l}}(\Omega_{r})\left[  Y_{l}^{m_{l}}(\Omega_{k})\right]  ^{\ast}%
\]
to write the coordinate-space projection of Eq.(\ref{sph-2-plane transform}):%
\[
4\pi C_{k}^{(uni)}%
{\displaystyle\sum_{l=0}^{\infty}}
i^{l}j_{l}(kr)%
{\displaystyle\sum_{\,m_{l}=-l}^{l}}
Y_{l}^{m_{l}}(\Omega_{r})\left[  Y_{l}^{m_{l}}(\Omega_{k})\right]  ^{\ast}=%
{\displaystyle\sum_{l=0}^{\infty}}
{\displaystyle\sum_{\,m_{l}=-l}^{l}}
U_{l}^{m_{l}}\left(  \widehat{k}\right)  C_{k}\,\,j_{l}(kr)Y_{l}^{m_{l}%
}(\Omega_{r})
\]
This will be satisfied if for every $l\geq0$ and $\left\vert m_{l}\right\vert
\leq l$ we have%
\[
4\pi C_{k}^{(uni)}i^{l}\left[  Y_{l}^{m_{l}}(\Omega_{k})\right]  ^{\ast}%
=U_{l}^{m_{l}}\left(  \widehat{k}\right)  C_{k}\,\,
\]
For $l$ and $m_{l}$ both zero this last relation reduces to $\sqrt{4\pi}%
C_{k}^{(uni)}=U_{0}^{0}\left(  \widehat{k}\right)  C_{k}$, leaving $\sqrt
{4\pi}U_{0}^{0}\left(  \widehat{k}\right)  i^{l}\left[  Y_{l}^{m_{l}}%
(\Omega_{k})\right]  ^{\ast}=U_{l}^{m_{l}}\left(  \widehat{k}\right)  $.
Setting $\sqrt{4\pi}U_{0}^{0}\left(  \widehat{k}\right)  =1$ then leads to%
\begin{equation}
U_{l}^{m_{l}}\left(  \widehat{k}\right)  =i^{l}\left[  Y_{l}^{m_{l}}%
(\Omega_{k})\right]  ^{\ast} \label{sph-uni map}%
\end{equation}
that describes the desired unitary transformation \cite{Arfken4}:%
\[%
{\displaystyle\sum_{l=0}^{\infty}}
{\displaystyle\sum_{\,m_{l}=-l}^{l}}
U_{l}^{m_{l}}\left(  \widehat{k_{2}}\right)  \left[  U_{l}^{m_{l}}\left(
\widehat{k_{1}}\right)  \right]  ^{\ast}=%
{\displaystyle\sum_{l=0}^{\infty}}
{\displaystyle\sum_{\,m_{l}=-l}^{l}}
\left[  Y_{l}^{m_{l}}(\Omega_{2})\right]  ^{\ast}Y_{l}^{m_{l}}(\Omega
_{1})=\delta\left(  \Omega_{1}\,-\,\Omega_{2}\right)
\]
It follows that the energy-normalized plane waves are described by the
normalizing factor%
\begin{equation}
C_{k}^{(uni)}=\frac{C_{k}}{\sqrt{4\pi}}U_{0}^{0}\left(  \widehat{k}\right)
=\frac{1}{4\pi}\sqrt{\frac{2mk}{\pi}} \label{norm-free3d}%
\end{equation}

\textit{Uni-directional time states} are formed from plane waves all moving in
the same direction, but with differing energy. Accordingly, we adopt the unit
vector $\widehat{k}$ as an additional label for such time states, writing
$|\,\tau\,\rangle\rightarrow|\,\tau,\widehat{k}\,\rangle$. These
uni-directional time states can be related to the angular momentum time states
of the preceding section. Combining Eqs.(\ref{history-semiopen}),
(\ref{sph-2-plane transform}), and (\ref{sph-uni map}), we find that the
uni-directional timeline wave in the coordinate basis, $\langle
\overrightarrow{r}\,|\,\tau,\widehat{k}\,\rangle\equiv\Xi_{\tau}%
^{\widehat{k}\,}(\overrightarrow{r})$, can be computed from the spherical-wave
expansion%
\begin{equation}
\Xi_{\tau}^{\widehat{k}\,}(\overrightarrow{r})=%
{\displaystyle\sum_{l=0}^{\infty}}
i^{l}\Xi_{\tau}^{l}(r)%
{\displaystyle\sum_{\,m_{l}=-l}^{l}}
Y_{l}^{m_{l}}(\Omega_{r})\left[  Y_{l}^{m_{l}}(\Omega_{k})\right]  ^{\ast},
\end{equation}
where $\Xi_{\tau}^{l}(r)$ is the radial timeline wave of
Eq.(\ref{waves-BesselJ}).\qquad

Alternatively, we might try to calculate $\Xi_{\tau}^{\widehat{k}%
\,}(\overrightarrow{r})$ directly by taking $|\,\psi\,\rangle
=|\,\overrightarrow{r}\,\rangle$ in Eq.(\ref{history-semiopen}). With the help
of Eqs.(\ref{plw-free3d}) and (\ref{norm-free3d}), we obtain in this way%
\begin{align}
\Xi_{\tau}^{\widehat{k}\,}(\overrightarrow{r})  &  =\frac{1}{\sqrt{2\pi}}%
{\displaystyle\int\limits_{0}^{\infty}}
\exp\left(  -iE_{k}\tau\right)  \,\langle\,\overrightarrow{r}%
\,|\,\overrightarrow{k}\,\rangle\,dE_{k}\nonumber\\
&  =\frac{1}{4\pi^{2}\sqrt{m}}%
{\displaystyle\int\limits_{0}^{\infty}}
\exp\left(  -ik^{2}\tau/2m+i\,k\widehat{k}\cdot\overrightarrow{r}\right)
\,k^{3/2}\,dk \label{integral-directional-3d}%
\end{align}
Eq.(\ref{integral-directional-3d}) shows that the dependence of $\Xi_{\tau
}^{\widehat{k}\,}(\overrightarrow{r})$ on $\widehat{k}$ and on
$\overrightarrow{r}$ occurs only through the combination $\xi\equiv
\widehat{k}\cdot\overrightarrow{r}$, which is nothing more than the projection
of the coordinate vector $\overrightarrow{r}$\ onto the direction of plane
wave propagation. (Indeed, $\Xi_{\tau}^{\widehat{k}\,}(\overrightarrow{r})$
itself is a plane wave -- albeit not a harmonic one -- with the surfaces of
constant wave amplitude oriented perpendicular to $\widehat{k}$.) In terms of
$\xi$, then, there is a universal timeline applicable to any direction in
space, as befits the expected isotropy of a free-particle environment. This
universal timeline has elements that we denote simply as $\Xi_{\tau}^{\,}%
(\xi)$, and are given by%
\begin{equation}
\Xi_{\tau}^{\,}(\xi)=\frac{1}{4\pi^{2}\sqrt{m}}%
{\displaystyle\int\limits_{0}^{\infty}}
\exp\left(  -ik^{2}\tau/2m+i\,k\xi\right)  \,k^{3/2}\,dk
\label{integral-universal}%
\end{equation}

Unlike a similar integral encountered in the one-dimensional case,
Eq.(\ref{integral-universal}) fails to converge for real values of $\tau$. But
the integral does define a function that is analytic throughout the lower half
plane $\Im m\,\tau<0$, and can be analytically continued onto the real axis.
For $\Re e\left(  \,i\tau\right)  >0$ we have \cite{Gradshteyn2}%
\begin{equation}
\Xi_{\tau}^{\,}(\xi)=\frac{3}{16\sqrt{\pi^{3}m}}z^{5/2}\exp\left(  -\xi
^{2}z^{2}/4\right)  D_{-5/2}\left(  -i\xi z\right)  \text{\qquad}z\equiv
\sqrt{\frac{m}{i\tau}}, \label{waves-directional-3d}%
\end{equation}
where $D_{-5/2}\left(  \ldots\right)  $ is another of the parabolic cylinder
functions. Eq.(\ref{waves-directional-3d}) limits $\tau$ to the sector
$-\pi<\arg\,\tau<0$ ($\Re e\left(  \,i\tau\right)  >0$), but the mapping from
$z$ to $\tau$ allows analytic continuation to the whole $\tau$-plane cut along
the negative real axis, $-\pi\leq\arg\,\tau<\pi$. The complex variable $z$
then is mapped into the sector $-3\pi/4<\arg\,z\leq\pi/4$. And because
$D_{\nu}$ is an entire function of its argument \cite{Bateman2},
Eq.(\ref{waves-directional-3d}) defines a single-valued function throughout
this domain. Comparing Eq.(\ref{waves-directional-3d}) for $\tau>0$
($\arg\,z=-\pi/4$) and $\tau<0$ ($\arg\,z=\pi/4$), we discover for all real
values of $\xi$ and $\tau$%
\begin{equation}
\Xi_{\tau}(\xi)=\left[  \Xi_{-\tau}(-\xi)\right]  ^{\ast}
\label{time-symmetry-univ}%
\end{equation}

For $\tau>0$ ($\arg\,z=-\pi/4$), the asymptotics of the parabolic cylinder
function \cite{Bateman3} imply%
\[
\Xi_{\tau}(\xi)\sim%
\begin{array}
[c]{cc}%
\exp\left(  im\xi^{2}/2\tau\right)  \xi^{3/2} & \qquad\xi>0\\
\left\vert \xi\right\vert ^{-5/2} & \qquad\xi<0
\end{array}
\]
Thus, $\Xi_{\tau}(\xi)$ diverges as $\xi^{3/2}$ for $\xi\rightarrow\infty$ and
vanishes as $\left\vert \xi\right\vert ^{-5/2}$ for $\xi\rightarrow-\infty$.
Analogous results for $\tau<0$ follow from Eq.(\ref{time-symmetry-univ}). The
behavior is reminiscent of the timeline functions constructed from running
waves in one dimension. Indeed, it appears that in $\Xi_{\tau}(\xi)$ we again
have time-of-arrival functions, with $-\tau$ denoting the arrival time at the
coordinate origin for waves moving in the direction of $\widehat{k}$, and
$\xi\equiv\widehat{k}\cdot\overrightarrow{r}$. This interpretation is
supported by the illustrations in Figs. 3 and 4 showing $\Xi_{\tau}(\xi)$ for
system times just prior to, and immediately following, arrival at the
coordinate origin.%
\begin{figure}[ptb]%
\centering
\includegraphics[
natheight=3.957400in,
natwidth=5.389500in,
height=2.4128in,
width=3.403in
]%
{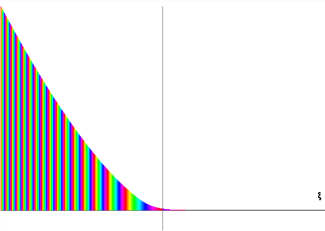}%
\caption{The universal timeline waveform $\Xi_{\tau}(\xi)$ for the system time
$\tau=-0.005$.\ The shading (coloring) represents varying phase values for
this complex function. In units where $\hbar=m=1$, the plot extends from
$\xi=-1$ to $\xi=+1$.}%
\label{fig3}%
\end{figure}
\begin{figure}[ptb]%
\centering
\includegraphics[
natheight=3.957400in,
natwidth=5.389500in,
height=2.4128in,
width=3.403in
]%
{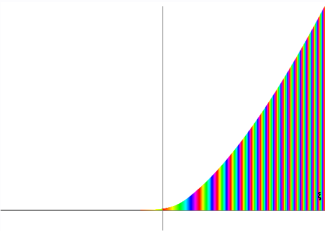}%
\caption{The universal timeline waveform $\Xi_{\tau}(\xi)$ for the system time
$\tau=+0.005$. The shading (coloring) represents varying phase values for this
complex function. In units where $\hbar=m=1$, the plot extends from $\xi=-1$
to $\xi=+1$.}%
\label{fig4}%
\end{figure}

\subsection{The Three-Dimensional Free-Particle Time Operator}

Lastly, we investigate the time operator for this example. We exercise the
freedom allowed by the degeneracy of free particle states to work in the
angular momentum representation. From Eq.(\ref{norm-spherical}) we find that
the stationary spherical waves $\langle\overrightarrow{r}\,|\,klm_{l}%
\,\rangle=C_{k}\,\,j_{l}(kr)Y_{l}^{m_{l}}(\Omega_{r})$ vanish at the lower
spectral edge ($\lim_{k\rightarrow0}\langle\overrightarrow{r}\,|\,klm_{l}%
\,\rangle=0$), so that a free-particle time operator in three space dimensions
does exist by the theory of Sec. III. The matrix elements of $\widehat{T}$ in
the coordinate basis are given by (cf. Eq.(\ref{time operator}))%
\begin{align*}
\langle\,\overrightarrow{r_{1}}\,|\widehat{T}_{3d-free}%
|\,\overrightarrow{r_{2}}\,\rangle &  =%
{\displaystyle\sum_{l,\,m_{l}}}
P%
{\displaystyle\int\limits_{-\infty}^{\infty}}
\langle\,\overrightarrow{r_{1}}\,|\,\tau lm_{l}\rangle\tau\langle\tau
lm_{l}\,|\,\overrightarrow{r_{2}}\,\rangle\,d\tau\\
&  =%
{\displaystyle\sum_{l,\,m_{l}}}
Y_{l}^{m_{l}}(\Omega_{1})\left[  Y_{l}^{m_{l}}(\Omega_{2})\right]  ^{\ast}P%
{\displaystyle\int\limits_{-\infty}^{\infty}}
\tau\,\Xi_{\tau}^{l}(r_{1})\left[  \Xi_{\tau}^{l}(r_{2})\right]  ^{\ast
}\,d\tau
\end{align*}
The principal value integrals are studied in Appendix A, where their existence
is rigorously established and a closed-form expression given for their
evaluation:%
\[
P%
{\displaystyle\int\limits_{-\infty}^{\infty}}
\tau\,\Xi_{\tau}^{l}(r_{1})\left[  \Xi_{\tau}^{l}(r_{2})\right]  ^{\ast
}\,d\tau=i\frac{m}{2}\operatorname{sgn}\left(  r_{1}-r_{2}\right)  \frac
{1}{r_{>}}\left(  \frac{r_{<}}{r_{>}}\right)  ^{l}%
\]
Collecting the above results, we obtain%
\[
\langle\,\overrightarrow{r_{1}}\,|\widehat{T}_{3d-free}%
|\,\overrightarrow{r_{2}}\,\rangle=i\frac{m}{2}\operatorname{sgn}\left(
r_{1}-r_{2}\right)  \frac{1}{r_{>}}%
{\displaystyle\sum_{l,\,m_{l}}}
Y_{l}^{m_{l}}(\Omega_{1})\left[  Y_{l}^{m_{l}}(\Omega_{2})\right]  ^{\ast
}\left(  \frac{r_{<}}{r_{>}}\right)  ^{l}%
\]

The remaining sums also can be evaluated in closed form. Combining the
generating function for the Legendre polynomials \cite{Arfken2} with the
addition theorem for spherical harmonics \cite{Arfken3}, we obtain for any
$\left\vert t\right\vert <1$%
\begin{align*}
\frac{1}{\sqrt{1-2t\cos\gamma+t^{2}}}  &  =%
{\displaystyle\sum_{l=0}^{\infty}}
P_{l}(\cos\gamma)t^{l}\\
&  =4\pi%
{\displaystyle\sum_{l=0}^{\infty}}
{\displaystyle\sum_{\,m_{l}=-l}^{l}}
Y_{l}^{m_{l}}(\Omega_{1})\left[  Y_{l}^{m_{l}}(\Omega_{2})\right]  ^{\ast
}\frac{t^{l}}{2l+1},
\end{align*}
from which it follows that%
\begin{align*}
2\pi%
{\displaystyle\sum_{l=0}^{\infty}}
{\displaystyle\sum_{\,m_{l}=-l}^{l}}
Y_{l}^{m_{l}}(\Omega_{1})\left[  Y_{l}^{m_{l}}(\Omega_{2})\right]  ^{\ast
}t^{l}  &  =t^{1/2}\frac{\partial}{\partial t}\left(  \frac{t^{1/2}}%
{\sqrt{1-2t\cos\gamma+t^{2}}}\right) \\
&  =\frac{1-t^{2}}{2\left(  1-2t\cos\gamma+t^{2}\right)  ^{3/2}}%
\end{align*}
Finally, identifying $t$ with $r_{<}/r_{>}$ gives%
\begin{align}
\langle\,\overrightarrow{r_{1}}\,|\widehat{T}_{3d-free}%
|\,\overrightarrow{r_{2}}\,\rangle &  =i\frac{m}{2}\operatorname{sgn}\left(
r_{1}-r_{2}\right)  \frac{1}{4\pi}\frac{r_{>}^{2}-r_{<}^{2}}{\left(  r_{>}%
^{2}-2r_{<}r_{>}\cos\gamma+r_{<}^{2}\right)  ^{3/2}}\nonumber\\
&  =\frac{im}{8\pi}\frac{\overrightarrow{r_{1}}\cdot\overrightarrow{r_{1}%
}-\overrightarrow{r_{2}}\cdot\overrightarrow{r_{2}}}{\,\left\vert
\overrightarrow{r_{1}}-\overrightarrow{r_{2}}\right\vert ^{3}}=\frac{im}{8\pi
}\frac{\left(  \overrightarrow{r_{1}}-\overrightarrow{r_{2}}\right)
}{\,\left\vert \overrightarrow{r_{1}}-\overrightarrow{r_{2}}\right\vert ^{3}%
}\cdot\left(  \overrightarrow{r_{1}}+\overrightarrow{r_{2}}\right)
\label{T-3d-free}%
\end{align}
The vector form for these matrix elements is pleasingly compact, and frees the
result from the spherical coordinates adopted for the computation.

It is apparent that Eq.(\ref{T-3d-free}) specifies matrix elements of a
Hermitian operator, i.e., $\langle\,\overrightarrow{r_{1}}\,|\widehat{T}%
_{3d-free}|\,\overrightarrow{r_{2}}\,\rangle=\langle\,\overrightarrow{r_{2}%
}\,|\widehat{T}_{3d-free}|\,\overrightarrow{r_{1}}\,\rangle^{\ast}$ for
$\overrightarrow{r_{1}}\neq\overrightarrow{r_{2}}$. That these matrix elements
also specify an operator that is canonically conjugate to the free-particle
Hamiltonian is confirmed in Appendix B. Thus, \textit{a canonical time
operator for a free particle in three dimensions exists, with coordinate-space
matrix elements given by Eq.(\ref{T-3d-free})}.

\section{Summary and Conclusions}

Contrary to conventional wisdom, we contend that [event] time is a legitimate
observable, and fits within the framework of standard quantum theory if we
extend the latter to include POVM's -- and not just self-adjoint operators --
for representing observables. This modest change in emphasis places the focus
squarely on probability amplitudes, in keeping with the seemingly evident fact
that [event] time statistics can be generated empirically for virtually any
quantum system. As with every other observable, we show these [event] time
statistics derive from wave functions expressed in a suitable basis (the time
basis), which is complete for the representation of any physical state. We
refer to this basis as a \textit{timeline}, or \textit{quantum history}, with
elements labeled by a continuous variable we call the \textit{system time}.
While time states are typically \textit{not orthogonal}, they do lead to wave
functions and statistics that are \textit{covariant} (time-translation
invariant), and probabilities that add to unity. We propose a recipe for
calculating wave functions in the time basis from those in the spectral basis.
This recipe is dictated solely by the demands of covariance and completeness,
and applies to virtually any Hamiltonian system. The phase ambiguity inherent
in the stationary states translates here into a freedom to construct time
statistics pertinent to different kinds of events.

The leap from time states to a time operator is non-trivial, involving
additional assumptions that are not always met. Indeed, it is the nature of
time statistics that they need not admit a well-defined mean, or variance.
Time operators -- when they exist -- are system specific, useful for
calculating moments of the [event] time distribution in those instances where
said moments can be shown to converge. Interestingly, we find that time
operators for periodic systems are never canonical to the Hamiltonian, but
canonical time operators can and do arise in [aperiodic] systems with a
vanishing point spectrum (no isolated levels).

As examples of these general principles, we have examined several systems
(particle in free-fall, free particle in one dimension) for which results have
been reported previously in the literature. Our objective has been to
illustrate how these diverse results follow from the unified approach
developed here. We also have gone beyond the familiar and applied that same
approach to the free particle in three dimensions. To the best of our
knowledge, results for the latter have never before appeared. Most
importantly, they confirm that the notion of an \textit{arrival time} -- first
encountered in the one-dimensional case -- extends to three dimensions,
complete with an accompanying canonical time operator. Possibilities for
future investigations abound. For instance, how to generate correct
arrival-time statistics for a particle scattering from even the simplest
one-dimensional barrier remains a subject of controversy \cite{Baute}; we
expect that discussion -- and numerous others -- to be informed by the results
presented here.

\appendix{}

\section{Time Operators and The Integrals $I_{l}\left(  r_{1},r_{2}\right)  $}

In this Appendix we investigate the principal value integrals that arise in
the construction of a canonical time operator for free particles:%
\begin{equation}
I_{l}\left(  r_{1},r_{2}\right)  \equiv P%
{\displaystyle\int\limits_{-\infty}^{\infty}}
\tau\,\Xi_{\tau}^{l}(r_{1})\left[  \Xi_{\tau}^{l}(r_{2})\right]  ^{\ast
}\,d\tau\equiv\lim_{\tau_{R}\rightarrow\infty}%
{\displaystyle\int\limits_{-\tau_{R}}^{+\tau_{R}}}
\tau\,\Xi_{\tau}^{l}(r_{1})\left[  \Xi_{\tau}^{l}(r_{2})\right]  ^{\ast
}\,d\tau
\end{equation}
Here $\Xi_{\tau}^{l}(r)$ is the timeline wavefunction in the $l^{th}$ angular
momentum subspace, given by Eq.(\ref{waves-BesselJ}). Inspection of
Eqs.(\ref{waves-free-o}) and (\ref{waves-free-e}) shows that the $l=-1$ and
$l=0$ integrals also appear in the context of the time operator for a free
particle in one dimension. Our objective here is to establish the existence of
these integrals, and obtain closed-form expressions suitable for their evaluation.

The relation $\Xi_{-\tau}^{l}(r)=\left[  \Xi_{\tau}^{l}(r)\right]  ^{\ast}$
can be used to show that $I_{l}\left(  r_{1},r_{2}\right)  $ is purely
imaginary, as well as antisymmetric under the interchange $r_{1}%
\longleftrightarrow r_{2}$, properties that can be used to reduce the integral
to the half-axis $\tau\geq0$:%
\begin{equation}
I_{l}\left(  r_{1},r_{2}\right)  =2i\,\Im m\left[  \lim_{\tau_{R}%
\rightarrow\infty}%
{\displaystyle\int\limits_{0}^{\tau_{R}}}
\tau\,\Xi_{\tau}^{l}(r_{1})\left[  \Xi_{\tau}^{l}(r_{2})\right]  ^{\ast
}\,d\tau\right]
\end{equation}
Substituting from Eq.(\ref{waves-BesselJ}), this becomes (recall
$2\alpha\equiv l-1/2$)%
\[
I_{l}\left(  r_{1},r_{2}\right)  =i\frac{m\sqrt{r_{1}r_{2}}}{2}\left(
I_{1}+I_{2}\right)
\]
where%

\begin{align*}
I_{1}  &  \equiv\lim_{s_{R}\rightarrow0^{+}}%
{\displaystyle\int\limits_{s_{R}}^{\infty}}
\sin\left(  s\left[  r_{1}^{2}-r_{2}^{2}\right]  \right)  \left[  J_{\alpha
+1}\left(  s\,r_{1}^{2}\right)  J_{\alpha+1}\left(  s\,r_{2}^{2}\right)
+J_{\alpha}\left(  s\,r_{1}^{2}\right)  J_{\alpha}\left(  s\,r_{2}^{2}\right)
\right]  \,ds\\
I_{2}  &  \equiv\lim_{s_{R}\rightarrow0^{+}}%
{\displaystyle\int\limits_{s_{R}}^{\infty}}
\cos\left(  s\left[  r_{1}^{2}-r_{2}^{2}\right]  \right)  \left[  J_{\alpha
+1}\left(  s\,r_{1}^{2}\right)  J_{\alpha}\left(  s\,r_{2}^{2}\right)
-J_{\alpha}\left(  s\,r_{1}^{2}\right)  J_{\alpha+1}\left(  s\,r_{2}%
^{2}\right)  \right]  \,ds
\end{align*}
The small-argument behavior of $J_{\alpha}$ ensures that both integrals exist
in the indicated limits provided $\alpha>-1$: accordingly, explicit reference
to the limits will be omitted from subsequent expressions.

Our next goal is to relate $I_{2}$ to $I_{1}$. To that end we define the
related (and simpler) integrals $\widetilde{I}_{1,2}$ by%
\begin{align*}
\widetilde{I}_{1}^{\alpha}\left(  r_{1},r_{2}\right)   &  \equiv%
{\displaystyle\int\limits_{0}^{\infty}}
\sin\left(  s\left[  r_{1}^{2}-r_{2}^{2}\right]  \right)  J_{\alpha}\left(
s\,r_{1}^{2}\right)  J_{\alpha}\left(  s\,r_{2}^{2}\right)  \,ds\\
\widetilde{I}_{2}^{\alpha}\left(  r_{1},r_{2}\right)   &  \equiv%
{\displaystyle\int\limits_{0}^{\infty}}
\cos\left(  s\left[  r_{1}^{2}-r_{2}^{2}\right]  \right)  J_{\alpha+1}\left(
s\,r_{1}^{2}\right)  J_{\alpha}\left(  s\,r_{2}^{2}\right)  \,ds
\end{align*}
Then $I_{1}\equiv\widetilde{I}_{1}^{\alpha+1}\left(  r_{1},r_{2}\right)
+\widetilde{I}_{1}^{\alpha}\left(  r_{1},r_{2}\right)  $ and $I_{2}%
\equiv\widetilde{I}_{2}^{\alpha}\left(  r_{1},r_{2}\right)  -\widetilde{I}%
_{2}^{\alpha}\left(  r_{2},r_{1}\right)  $. Integrating $\widetilde{I}_{2}$
once by parts (the out-integrated part vanishes for $\alpha>-1$) and using the
Bessel recursion relation \cite{Bateman}%
\[
J_{\nu-1}\left(  z\right)  -J_{\nu+1}\left(  z\right)  =2\frac{dJ_{\nu}\left(
z\right)  }{dz}%
\]
results in%
\begin{align*}
\widetilde{I}_{2}^{\alpha}\left(  r_{1},r_{2}\right)   &  =-\frac{1}{r_{1}%
^{2}-r_{2}^{2}}%
{\displaystyle\int\limits_{0}^{\infty}}
\sin\left(  s\left[  r_{1}^{2}-r_{2}^{2}\right]  \right)  \frac{\partial
}{\partial s}\left[  J_{\alpha+1}\left(  s\,r_{1}^{2}\right)  J_{\alpha
}\left(  s\,r_{2}^{2}\right)  \right]  \,ds\\
&  =\frac{r_{2}^{2}}{2\left(  r_{1}^{2}-r_{2}^{2}\right)  }%
{\displaystyle\int\limits_{0}^{\infty}}
\sin\left(  s\left[  r_{1}^{2}-r_{2}^{2}\right]  \right)  J_{\alpha+1}\left(
s\,r_{1}^{2}\right)  \left[  J_{\alpha+1}\left(  s\,r_{2}^{2}\right)
-J_{\alpha-1}\left(  s\,r_{2}^{2}\right)  \right]  \,ds\\
&  +\frac{r_{1}^{2}}{2\left(  r_{1}^{2}-r_{2}^{2}\right)  }%
{\displaystyle\int\limits_{0}^{\infty}}
\sin\left(  s\left[  r_{1}^{2}-r_{2}^{2}\right]  \right)  J_{\alpha}\left(
s\,r_{2}^{2}\right)  \left[  J_{\alpha+2}\left(  s\,r_{1}^{2}\right)
-J_{\alpha}\left(  s\,r_{1}^{2}\right)  \right]  \,ds
\end{align*}
In terms of yet a third integral $\widetilde{I}_{3}$ defined as%
\begin{align*}
\widetilde{I}_{3}^{\alpha}\left(  r_{1},r_{2}\right)   &  \equiv%
{\displaystyle\int\limits_{0}^{\infty}}
\sin\left(  s\left[  r_{1}^{2}-r_{2}^{2}\right]  \right)  J_{\alpha+1}\left(
s\,r_{1}^{2}\right)  J_{\alpha-1}\left(  s\,r_{2}^{2}\right)  \,ds\\
&  =\frac{2\alpha}{r_{2}^{2}}%
{\displaystyle\int\limits_{0}^{\infty}}
\sin\left(  s\left[  r_{1}^{2}-r_{2}^{2}\right]  \right)  J_{\alpha+1}\left(
s\,r_{1}^{2}\right)  J_{\alpha}\left(  s\,r_{2}^{2}\right)  \,\frac{ds}%
{s}-\widetilde{I}_{1}^{\alpha+1}\left(  r_{1},r_{2}\right) \\
&  =\frac{2\alpha}{r_{1}^{2}}%
{\displaystyle\int\limits_{0}^{\infty}}
\sin\left(  s\left[  r_{1}^{2}-r_{2}^{2}\right]  \right)  J_{\alpha}\left(
s\,r_{1}^{2}\right)  J_{\alpha-1}\left(  s\,r_{2}^{2}\right)  \,\frac{ds}%
{s}-\widetilde{I}_{1}^{\alpha-1}\left(  r_{1},r_{2}\right)
\end{align*}
we can write the result for $\widetilde{I}_{2}$ compactly as%
\begin{align*}
\widetilde{I}_{2}^{\alpha}\left(  r_{1},r_{2}\right)   &  =\frac{1}{2\left(
r_{1}^{2}-r_{2}^{2}\right)  }\left[  -r_{1}^{2}\widetilde{I}_{1}^{\alpha
}\left(  r_{1},r_{2}\right)  +r_{2}^{2}\widetilde{I}_{1}^{\alpha+1}\left(
r_{1},r_{2}\right)  -r_{2}^{2}\widetilde{I}_{3}^{\alpha}\left(  r_{1}%
,r_{2}\right)  +r_{1}^{2}\widetilde{I}_{3}^{\alpha+1}\left(  r_{1}%
,r_{2}\right)  \right] \\
&  =\frac{1}{\left(  r_{1}^{2}-r_{2}^{2}\right)  }\left[  -r_{1}%
^{2}\widetilde{I}_{1}^{\alpha}\left(  r_{1},r_{2}\right)  +r_{2}%
^{2}\widetilde{I}_{1}^{\alpha+1}\left(  r_{1},r_{2}\right)  +%
{\displaystyle\int\limits_{0}^{\infty}}
\sin\left(  s\left[  r_{1}^{2}-r_{2}^{2}\right]  \right)  J_{\alpha+1}\left(
s\,r_{1}^{2}\right)  J_{\alpha}\left(  s\,r_{2}^{2}\right)  \,\frac{ds}%
{s}\right]
\end{align*}
Then%
\begin{align*}
I_{2}  &  =\frac{1}{\left(  r_{1}^{2}-r_{2}^{2}\right)  }\left[  -r_{1}%
^{2}\widetilde{I}_{1}^{\alpha}\left(  r_{1},r_{2}\right)  +r_{2}%
^{2}\widetilde{I}_{1}^{\alpha+1}\left(  r_{1},r_{2}\right)  +r_{2}%
^{2}\widetilde{I}_{1}^{\alpha}\left(  r_{1},r_{2}\right)  -r_{1}%
^{2}\widetilde{I}_{1}^{\alpha+1}\left(  r_{1},r_{2}\right)  \right] \\
&  +\frac{1}{\left(  r_{1}^{2}-r_{2}^{2}\right)  }%
{\displaystyle\int\limits_{0}^{\infty}}
\sin\left(  s\left[  r_{1}^{2}-r_{2}^{2}\right]  \right)  \left[  J_{\alpha
+1}\left(  s\,r_{1}^{2}\right)  J_{\alpha}\left(  s\,r_{2}^{2}\right)
-J_{\alpha+1}\left(  s\,r_{2}^{2}\right)  J_{\alpha}\left(  s\,r_{1}%
^{2}\right)  \right]  \,\frac{ds}{s}\\
&  =-I_{1}+\frac{1}{\left(  r_{1}^{2}-r_{2}^{2}\right)  }%
{\displaystyle\int\limits_{0}^{\infty}}
\sin\left(  s\left[  r_{1}^{2}-r_{2}^{2}\right]  \right)  \left[  J_{\alpha
+1}\left(  s\,r_{1}^{2}\right)  J_{\alpha}\left(  s\,r_{2}^{2}\right)
-J_{\alpha+1}\left(  s\,r_{2}^{2}\right)  J_{\alpha}\left(  s\,r_{1}%
^{2}\right)  \right]  \,\frac{ds}{s}%
\end{align*}
and so%
\begin{equation}
I_{l}\left(  r_{1},r_{2}\right)  =i\frac{m\sqrt{r_{1}r_{2}}}{2\left(
r_{1}^{2}-r_{2}^{2}\right)  }%
{\displaystyle\int\limits_{0}^{\infty}}
\sin\left(  s\left[  r_{1}^{2}-r_{2}^{2}\right]  \right)  \left[  J_{\alpha
+1}\left(  s\,r_{1}^{2}\right)  J_{\alpha}\left(  s\,r_{2}^{2}\right)
-J_{\alpha+1}\left(  s\,r_{2}^{2}\right)  J_{\alpha}\left(  s\,r_{1}%
^{2}\right)  \right]  \,\frac{ds}{s} \label{integral-reduced}%
\end{equation}

Finally, from integral tables \cite{Gradshteyn3} we have for $a,b>0$ and
$\alpha>-1$:%
\begin{align*}%
{\displaystyle\int\limits_{0}^{\infty}}
J_{\alpha+1}\left(  ax\right)  J_{\alpha}\left(  bx\right)  \sin\left(
cx\right)  \,\frac{dx}{x}  &  =0\text{ \ \ \ for }0<c<b-a\\
&  =a^{-\alpha-1}b^{\alpha}c\text{ \ \ for }0<c<a-b
\end{align*}
This result is applied separately to each integral in
Eq.(\ref{integral-reduced}). In terms of the smaller ($r_{<}$) and larger
($r_{>}$) of its two arguments, the final form for $I_{l}\left(  r_{1}%
,r_{2}\right)  $ can be written most compactly as%
\begin{equation}
I_{l}\left(  r_{1},r_{2}\right)  =i\frac{m}{2}\operatorname{sgn}\left(
r_{1}-r_{2}\right)  \frac{1}{r_{>}}\left(  \frac{r_{<}}{r_{>}}\right)  ^{l}
\label{integral-result}%
\end{equation}

\section{Canonical Property of the Free-Particle Time Operator in Three
Dimensions}

In this Appendix we establish that the time operator of Eq.(\ref{T-3d-free})
is canonical to the free-particle Hamiltonian $\widehat{H}=\widehat{p}^{2}%
/2m$. To that end we examine the coordinate-space matrix elements of the
commutator%
\begin{align}
\langle\,\overrightarrow{r_{1}}\,|\left[  \widehat{T}_{3d-free},\widehat{H}%
\right]  |\,\overrightarrow{r_{2}}\,\rangle &  =\frac{1}{2m}\langle
\,\overrightarrow{r_{1}}\,|\widehat{T}_{3d-free}\widehat{p}^{2}-\widehat{p}%
^{2}\widehat{T}_{3d-free}|\,\overrightarrow{r_{2}}\,\rangle\nonumber\\
&  =\frac{1}{2m}\left(  \nabla_{1}^{2}-\nabla_{2}^{2}\right)  \langle
\,\overrightarrow{r_{1}}\,|\widehat{T}_{3d-free}|\,\overrightarrow{r_{2}%
}\,\rangle\label{laplacian12}%
\end{align}
For evaluating the Laplacians in this expression, we apply the vector calculus
identity \cite{Griffiths}%
\begin{equation}
\nabla\left(  \overrightarrow{A}\cdot\overrightarrow{B}\right)
=\overrightarrow{A}\times\left(  \nabla\times\overrightarrow{B}\right)
+\overrightarrow{B}\times\left(  \nabla\times\overrightarrow{A}\right)
+\left(  \overrightarrow{A}\cdot\nabla\right)  \overrightarrow{B}+\left(
\overrightarrow{B}\cdot\nabla\right)  \overrightarrow{A} \label{grad-dot}%
\end{equation}
with the identifications%
\[
\overrightarrow{A}=\frac{\left(  \overrightarrow{r_{1}}-\overrightarrow{r_{2}%
}\right)  }{\,\left\vert \overrightarrow{r_{1}}-\overrightarrow{r_{2}%
}\right\vert ^{3}};\quad\overrightarrow{B}=\overrightarrow{r_{1}%
}+\overrightarrow{r_{2}}%
\]
Noting that $\overrightarrow{A}$ is essentially the electrostatic field of a
point charge, we have%
\begin{align*}
\nabla_{1}\cdot\overrightarrow{A}  &  =4\pi\,\delta\left(
\overrightarrow{r_{1}}-\overrightarrow{r_{2}}\right)  =-\nabla_{2}%
\cdot\overrightarrow{A}\\
\nabla_{1}\times\overrightarrow{A}  &  =0=\nabla_{2}\times\overrightarrow{A}%
\end{align*}
$\overrightarrow{B}$ also is curl-free, so the first two terms in
Eq.(\ref{grad-dot}) are zero. Further, with $\overrightarrow{r_{1,2}}=\left(
x_{1,2},\,y_{1,2},\,z_{1,2}\right)  $, the simplicity of $\overrightarrow{B}$
allows us to write%
\begin{align*}
\left(  \overrightarrow{A}\cdot\nabla_{1}\right)  B_{x}  &  =\left(
A_{x}\frac{\partial}{\partial x_{1}}+A_{y}\frac{\partial}{\partial y_{1}%
}+A_{z}\frac{\partial}{\partial z_{1}}\right)  \left(  x_{1}+x_{2}\right) \\
&  =A_{x},\text{ \ etc.}%
\end{align*}
and%
\begin{align*}
\frac{\partial}{\partial x_{1}}\left[  \left(  \overrightarrow{B}\cdot
\nabla_{1}\right)  A_{x}\right]   &  =\frac{\partial}{\partial x_{1}}\left[
\left(  x_{1}+x_{2}\right)  \frac{\partial A_{x}}{\partial x_{1}}+\left(
y_{1}+y_{2}\right)  \frac{\partial A_{x}}{\partial y_{1}}+\left(  z_{1}%
+z_{2}\right)  \frac{\partial A_{x}}{\partial z_{1}}\right] \\
&  =\left(  \overrightarrow{B}\cdot\nabla_{1}\right)  \frac{\partial A_{x}%
}{\partial x_{1}}+\frac{\partial A_{x}}{\partial x_{1}},\text{ \ etc.}%
\end{align*}
Then%
\begin{equation}
\nabla_{1}^{2}\left(  \overrightarrow{A}\cdot\overrightarrow{B}\right)
=\nabla_{1}\cdot\nabla_{1}\left(  \overrightarrow{A}\cdot\overrightarrow{B}%
\right)  =2\left(  \nabla_{1}\cdot\overrightarrow{A}\right)  +\left(
\overrightarrow{B}\cdot\nabla_{1}\right)  \left(  \nabla_{1}\cdot
\overrightarrow{A}\right)  \label{laplacian1}%
\end{equation}
Replacing $\nabla_{1}$ with $\nabla_{2}$ in this last expression generates an
equally valid result, but since $\overrightarrow{A}$ depends only on
$\overrightarrow{r_{1}}-\overrightarrow{r_{2}}$, we obtain%
\begin{equation}
\nabla_{2}^{2}\left(  \overrightarrow{A}\cdot\overrightarrow{B}\right)
=-2\left(  \nabla_{1}\cdot\overrightarrow{A}\right)  +\left(
\overrightarrow{B}\cdot\nabla_{1}\right)  \left(  \nabla_{1}\cdot
\overrightarrow{A}\right)  \label{laplacian2}%
\end{equation}
and finally,%
\begin{align}
\langle\,\overrightarrow{r_{1}}\,|\left[  \widehat{T}_{3d-free},\widehat{H}%
\right]  |\,\overrightarrow{r_{2}}\,\rangle &  =i\frac{1}{16\pi}\left(
\nabla_{1}^{2}-\nabla_{2}^{2}\right)  \left(  \overrightarrow{A}%
\cdot\overrightarrow{B}\right) \nonumber\\
&  =i\frac{1}{4\pi}\,\left(  \nabla_{1}\cdot\overrightarrow{A}\right)
=i\,\delta\left(  \overrightarrow{r_{1}}-\overrightarrow{r_{2}}\right)
\label{commutator-3d-free}%
\end{align}
We conclude that $\left[  \widehat{T}_{3d-free},\widehat{H}\right]  =i$, i.e.,
that $\widehat{T}_{3d-free}$ is canonical to the free-particle Hamiltonian
$\widehat{H}$.


\begin{thebibliography}{99}
\bibitem {eMax}We expect that $E_{\max}$ grows without bound as $N$ increases,
implying that there is no natural upper limit to the energy spectrum. While
this may seem evident on its face, there exist model spectra for which it is
not true (e.g., the discrete spectrum of atomic hydrogen).
\end{thebibliography}
\end{document}